\documentclass[times,twocolumn,final,longtitle]{elsarticle}

\usepackage{medima}
\usepackage{framed,multirow}

\usepackage{mwe}
\usepackage{blindtext}

\usepackage{amssymb}

\usepackage{url}
\usepackage{xcolor,colortbl}
\usepackage{amsmath}
\usepackage{tikz}
\usepackage{booktabs}

\usepackage{subfig} 

\usepackage{rotating} %

\definecolor{Samoyed-c}{rgb}{0.86,0.371,0.34}
\definecolor{PKU-BIALAB-c}{rgb}{0.86,0.566,0.34}
\definecolor{jwc-rad-c}{rgb}{0.86,0.761,0.34}
\definecolor{MIP-c}{rgb}{0.764,0.86,0.34}
\definecolor{PremiLab-c}{rgb}{0.569,0.86,0.34}
\definecolor{Epione-Liryc-c}{rgb}{0.374,0.86,0.34}
\definecolor{MedICL-c}{rgb}{0.34,0.86,0.501}
\definecolor{DBMI-pitt-c}{rgb}{0.34,0.86,0.696}
\definecolor{Hi-Lib-c}{rgb}{0.34,0.829,0.86}
\definecolor{smriti161096-c}{rgb}{0.34,0.634,0.86}
\definecolor{IMI-c}{rgb}{0.34,0.439,0.86}
\definecolor{GapMIND-c}{rgb}{0.436,0.34,0.86}
\definecolor{gabybaldeon-c}{rgb}{0.631,0.34,0.86}
\definecolor{SEU-Chen-c}{rgb}{0.826,0.34,0.86}
\definecolor{skjp-c}{rgb}{0.86,0.34,0.699}
\definecolor{IRA-c}{rgb}{0.86,0.34,0.504}

\definecolor{Gray}{gray}{0.85}
\definecolor{LightGray}{gray}{0.95}

\DeclareRobustCommand\circle[1]{\setlength{\fboxrule}{0pt}%
\fbox{\tikz\draw[#1,fill=#1] (0,0) circle (.7ex);}}

\usepackage{xspace}
\newcommand\Tone{ceT\textsubscript{1}\xspace}
\newcommand\Ttwo{hrT\textsubscript{2}\xspace}

\DeclareMathOperator{\DSC}{DSC}
\DeclareMathOperator{\ASSD}{ASSD}

\usepackage{hyperref}

\journal{Submitted to Medical Image Analysis}

\begin{document}

\verso{Reuben Dorent \textit{et~al.}}

\begin{frontmatter}

\title{CrossMoDA 2021 challenge: Benchmark of Cross-Modality Domain Adaptation techniques for Vestibular Schwannoma and Cochlea Segmentation}%

\author[1]{Reuben \snm{Dorent}\texorpdfstring{\corref{cor1}}{}}
\cortext[cor1]{Corresponding author}
\ead{reuben.dorent@kcl.ac.uk}
\author[1]{Aaron \snm{Kujawa}}
\author[1]{Marina \snm{Ivory}}
\author[2,23,24]{Spyridon \snm{Bakas}}
\author[3]{Nicola \snm{Rieke}}
\author[1]{Samuel \snm{Joutard}}
\author[4]{Ben \snm{Glocker}}
\author[1]{Jorge \snm{Cardoso}}
\author[1]{Marc \snm{Modat}}

\author[12]{Kayhan \snm{Batmanghelich}}
\author[19]{Arseniy \snm{Belkov}}
\author[16]{Maria \snm{ Baldeon Calisto}}
\author[7]{Jae Won \snm{Choi}}
\author[8]{Benoit M. \snm{Dawant}}
\author[6]{Hexin \snm{Dong}}
\author[14]{Sergio \snm{Escalera}}
\author[8]{Yubo \snm{Fan}}
\author[15]{Lasse \snm{Hansen}}
\author[15]{Mattias P. \snm{Heinrich}}
\author[14]{Smriti \snm{Joshi}}
\author[11]{Victoriya \snm{Kashtanova}}
\author[5]{Hyeon Gyu  \snm{Kim}}
\author[18]{Satoshi \snm{Kondo}}
\author[15]{Christian N. \snm{Kruse}}
\author[17]{Susana K. \snm{Lai-Yuen}}
\author[8]{Hao  \snm{Li}}
\author[8]{Han \snm{Liu}}
\author[11]{Buntheng \snm{Ly}}
\author[8]{Ipek \snm{Oguz}}
\author[5]{Hyungseob \snm{Shin}}
\author[20,21]{Boris \snm{Shirokikh}}
\author[9,10]{Zixian \snm{Su}}
\author[13]{Guotai \snm{Wang}}
\author[13]{Jianghao \snm{Wu}}
\author[12]{Yanwu \snm{Xu}}
\author[9,10]{Kai \snm{Yao}}
\author[6]{Li  \snm{Zhang}}

\author[1]{Sébastien \snm{Ourselin}}
\author[1,22]{Jonathan \snm{Shapey}}
\author[1]{Tom \snm{Vercauteren}}


\address[1]{School of Biomedical Engineering $\&$ Imaging Sciences, King's College London, London, United Kingdom}
\address[2]{Center for Biomedical Image Computing and Analytics (CBICA), University of Pennsylvania, Philadelphia, USA}
\address[23]{Department of Pathology and Laboratory Medicine, Perelman School of Medicine, University of Pennsylvania, Philadelphia, PA, USA}
\address[24]{Department of Radiology, Perelman School of Medicine, University of Pennsylvania, Philadelphia, PA, USA}
\address[3]{NVIDIA}
\address[4]{Department of Computing, Imperial College London, Department of Computing, London, United Kingdom}
\address[5]{School of Electrical and Electronic Engineering, Yonsei University, Seoul, Korea}
\address[6]{Center for Data Science, Peking University, Beijing, China}
\address[7]{Department of Radiology, Armed Forces Yangju Hospital, Yangju, Korea}
\address[8]{Vanderbilt University, Nashville, USA}
\address[9]{University of Liverpool, Liverpool, United Kingdom}
\address[10]{School of Advanced Technology, Xi'an Jiaotong-Liverpool University, Suzhou,  China}
\address[11]{Inria, Université Côte d’Azur, Sophia Antipolis, France}
\address[12]{Department of Biomedical Informatics, University of Pittsburgh, Pittsburgh, USA}
\address[13]{School of Mechanical and Electrical Engineering,
University of Electronic Science and Technology of China, Chengdu, China}
\address[14]{Artificial Intelligence in Medicine Lab (BCN-AIM) and Human Behavior Analysis Lab (HuPBA), Universitat de Barcelona, Barcelona, Spain}
\address[15]{Institute of Medical Informatics, Universität zu Lübeck, Germany}
\address[16]{Universidad San Francisco de Quito, Quito, Ecuador}
\address[17]{University of South Florida, Tampa, USA}
\address[18]{Muroran Institute of Technology, Muroran, Japan}
\address[19]{Moscow Institute of Physics and Technology, Moscow, Russia}
\address[20]{Skolkovo Institute of Science and Technology, Moscow, Russia}
\address[21]{Artificial Intelligence Research Institute (AIRI), Moscow, Russia}
\address[22]{Department of Neurosurgery, King's College Hospital, London, United Kingdom}



\begin{abstract}
Domain Adaptation (DA) has recently been of strong interest in the medical imaging community.
While a large variety of DA techniques have been proposed for image segmentation, most of these techniques have been validated either on private datasets or on small publicly available datasets. Moreover, these datasets mostly addressed single-class problems. 

To tackle these limitations, the Cross-Modality Domain Adaptation (crossMoDA) challenge was organised in conjunction with the 24th International Conference on Medical Image Computing and Computer Assisted Intervention (MICCAI 2021). CrossMoDA is the first large and multi-class benchmark for unsupervised cross-modality Domain Adaptation. The goal of the challenge is to segment two key brain structures involved in the follow-up and treatment planning of vestibular schwannoma (VS): the VS and the cochleas. Currently, the diagnosis and surveillance in patients with VS are commonly performed using contrast-enhanced T1 (\Tone) MR imaging. However, there is growing interest in using non-contrast imaging sequences such as high-resolution T2 (\Ttwo) imaging. 
For this reason, we established an unsupervised cross-modality segmentation benchmark.
The training dataset provides annotated \Tone scans (N=105) and unpaired non-annotated \Ttwo scans (N=105).
The aim was to automatically perform unilateral VS and bilateral cochlea segmentation on \Ttwo scans as provided in the testing set (N=137).
This problem is particularly challenging given the large intensity distribution gap across the modalities and the small volume of the structures.  


A total of 55 teams from 16 countries submitted predictions to the validation leaderboard. Among them, 16 teams from 9 different countries submitted their algorithm for the evaluation phase. The level of performance reached by the top-performing teams is
strikingly
high (best median Dice score - VS: $88.4\%$; Cochleas: $85.7\%$) and close to full supervision (median Dice score - VS: $92.5\%$; Cochleas: $87.7\%$).
All top-performing methods made use of
an image-to-image translation
approach to transform the source-domain images into pseudo-target-domain images. A segmentation network was then trained using these generated images and the manual annotations provided for the source image.

\end{abstract}


\end{frontmatter}




\begin{figure*}[tb!]
  \includegraphics[width=\linewidth]{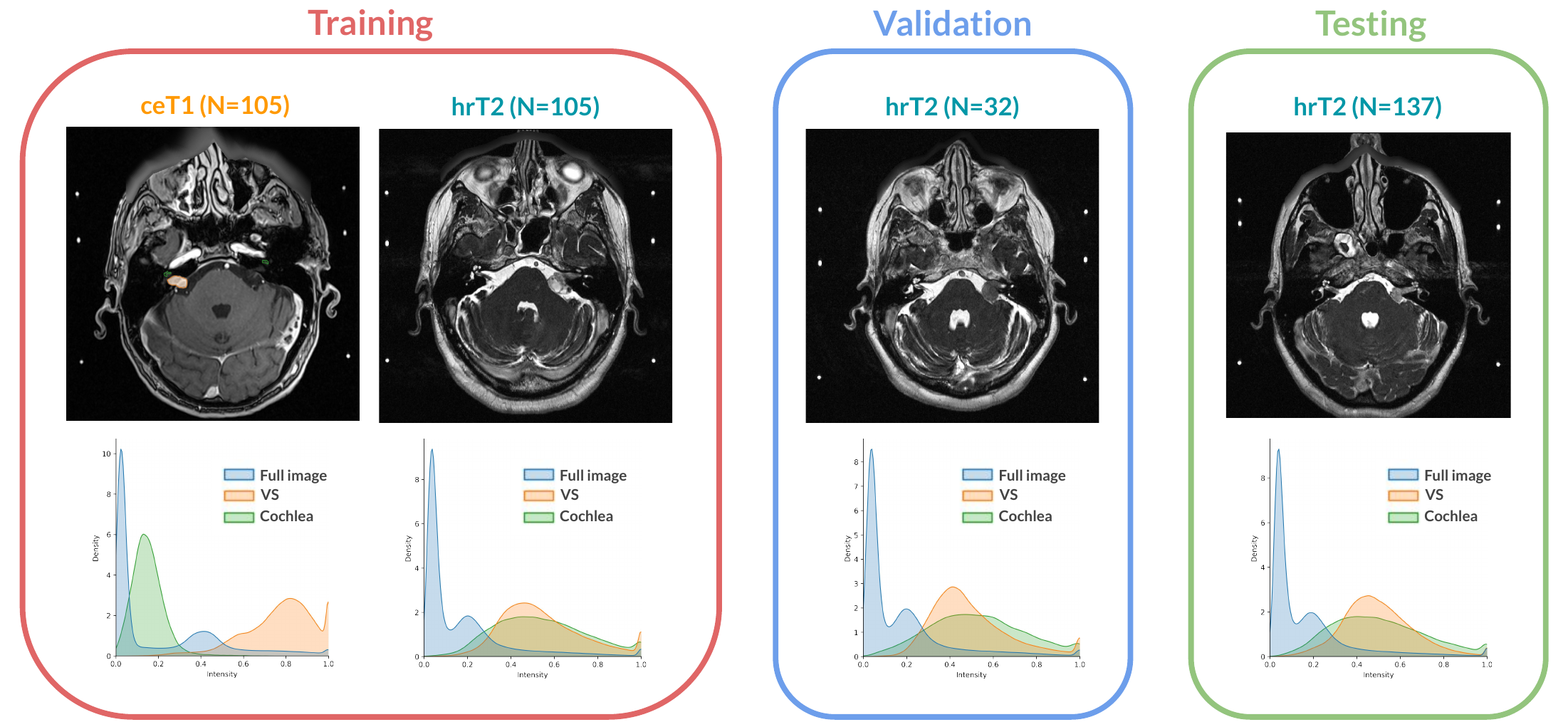}
  \caption{Overview of the challenge dataset. Annotations are only available for the training \Tone scans. Intensity distribution on each set are shown per structure. The intensity is normalised between 0 and 1 for each volume.}
  \label{fig:data_set_pres}
\end{figure*}

\section{Introduction}

Machine learning (ML) has recently reached outstanding performance in medical image analysis. These techniques typically assume that the training dataset (source domain) and test dataset (target domain) are drawn from the same data distribution. However, this assumption does not always stand in clinical practice. For example, the data may have been acquired at different medical centres, with different scanners, and under different image acquisition protocols. Recent studies have shown that ML algorithms, including deep learning ones, are particularly sensitive to data changes and experience performance drops due to domain shifts \citep{6945865,pmlr-v32-donahue14}. This domain shift problem strongly reduces the applicability of ML approaches to real-world clinical settings.

To increase the robustness of ML techniques, a naive approach aims at training models on large-scale datasets that cover large data variability. Therefore, efforts have been made in the computer vision community to collect and annotate data. For example, the Open Images dataset \citep{OpenImages} contains 9 million varied images with rich annotations. While natural images can be easily collected from the Internet, access to medical data is often restricted to preserve medical privacy. Moreover, annotating medical images is time-consuming and expensive as it requires the expertise of physicians, radiologists, and surgeons. For these reasons, it is unlikely that large, annotated and open databases will become available for most medical problems.

To address the lack of large amounts of labelled medical data, domain adaptation (DA) has  been of strong interest in the medical imaging community.
DA is a subcategory of transfer learning that aims at bridging the domain distribution discrepancy between the source domain and the target domain.
While the source and target data are assumed to be available at training time, target label availability is either limited (supervised and semi-supervised DA), incomplete (weakly-supervised DA) or missing (unsupervised DA). A complete review of DA for medical image analysis can be found in \cite{surveyDA}.
Unsupervised DA (UDA) has especially raised attention as it doesn't require any additional annotations. However, existing UDA techniques have been either tested on private, small or single class datasets. Consequently, there is a need for a public benchmark on a large and multi-class dataset. 

To benchmark new and existing unsupervised DA techniques for medical image segmentation, we organised the crossMoDA challenge in conjunction with the 24th International Conference on Medical Image Computing and Computer Assisted Intervention (MICCAI 2021). The goal of the challenge was to segment two key brain structures involved in the follow-up and treatment planning of vestibular schwannoma (VS): the VS and the cochleas. With data from 379 patients, crossMoDA is the first large and multi-class benchmark for unsupervised cross-modality domain adaptation.

VS is a benign tumour arising from the nerve sheath of one of the vestibular nerves. The incidence of VS has been estimated to be 1 in 1000 \citep{00129492-200501000-00016}. For smaller tumours, observation using MR imaging is often advised. If the tumour demonstrates growth, management options include conventional open surgery or stereotactic radiosurgery (SRS), which requires the segmentation of VS and the surrounding organs at risk (e.g., the cochlea) \citep{SHAPEY2021269}. The tumour's maximal linear dimension is typically measured to estimate the tumour growth. However, recent studies \citep{MacKeith2018, Growthofuntreatedvestibularschwannomaaprospectivestudy} have demonstrated that a volumetric measurement is a more accurate and sensitive method of calculating a VS's true size and is superior at detecting subtle growth. For these reasons, automated methods for VS delineation have been recently proposed~\citep{10.1007/978-3-030-32245-8_96,shapey2019artificial,Lee2021, 10.1007/978-3-030-87196-3_57}.

Currently, the diagnosis and surveillance of patients with VS are commonly performed using contrast-enhanced T1 (\Tone) MR imaging. However, there is growing interest in using non-contrast imaging sequences such as high-resolution T2 (\Ttwo) imaging, as it mitigates the risks associated with gadolinium-containing contrast agents \citep{Khawaja2015}. In addition to improving patient safety, \Ttwo imaging is 10 times more cost-efficient than  \Tone imaging \citep{Coelho2018}. For this reason, we proposed a cross-modality benchmark (from \Tone to \Ttwo) that aims to automatically perform VS and cochleas segmentation on \Ttwo scans.

This paper summarises the 2021 challenge and is structured as follows. First, a review of existing datasets used to assess existing domain adaptation techniques for image segmentation is proposed in Section~\ref{sec:related_work}. Then, the design of the crossMoDA challenge is given in Section~\ref{sec:challenge_description}. Section~\ref{sec:evaluation} presents the evaluation strategy of the challenge (metrics and ranking scheme). Participating methods are then described and compared in Section~\ref{sec:participating_methods}. Finally, Section~\ref{sec:results} presents the results obtained by the participating team and Section~\ref{sec:discussion} provides a discussion and concludes the paper.


\section{Related work}\label{sec:related_work}
We performed a literature review to survey the benchmark datasets used to assess DA techniques for unsupervised medical image segmentation.
On the methodological side, as detailed afterwards, a range of methods was used by the participating teams, illustrating a wide breadth of different DA approaches.
Nonetheless, a thorough review of DA methodologies is out of the scope of this paper. We refer
the interested reader to \cite{surveyDA} for a recent review of these.

Many domain adaptation techniques for medical image segmentation have been validated on private datasets, for example \cite{10.1007/978-3-319-59050-9_47,10.1007/978-3-030-32245-8_29}. Given that these datasets used for the experiments are not publicly available, it is not possible to compare new methods with these techniques. 

Other authors have used public datasets to validate their methods. Interestingly, these datasets often come from previous medical segmentation challenges that weren't originally proposed for domain adaptation. For this reason, unsupervised problems are generated by artificially removing annotations on subsets of these challenge datasets.
We 
present these
open
datasets and highlight their limitations for evaluating unsupervised domain adaptation:
\begin{itemize}
    \item \textit{WMH}: The MICCAI White Matter Hyperintensities (WMH) Challenge dataset \citep{WMH} consists of brain MR images with manual annotations of WMH from three different institutions. Each institution provided 20 multi-modal images for the training set. Domain adaptation techniques have been validated on each set of scans acquired at the same institution \citep{orbes2019multi,10.1117/12.2579548,SUNDARESAN2021102215}. Each institution set ($N=20$) is not only used to assess the methods but also to perform domain adaptation during training. Consequently, the test sets are extremely small ($N\leq10$ scans), leading to comparisons with low statistical power. Another limitation of this dataset is that it only assesses single-class UDA solutions. Finally, the domain shift is limited as the source and target domains correspond to the same image modalities acquired with 3T MRI scanners.
    
    \item \textit{SCGM}: The Spinal Cord Gray Matter Challenge (SCGM) dataset is a collection of cervical MRI from four institutions \citep{SCGM}. Each site provided unimodal images from 20 healthy subjects along with manual segmentation masks. Various unsupervised domain adaptation techniques have been tested on this dataset \citep{PERONE20191,LIU2021102214,10.1007/978-3-030-33391-1_4}. Again, the main limitation of this dataset is the small size of the test sets ($N\leq20$ scans). Moreover, the problem is single-class, and the domain shift is limited (intra-modality UDA).
    
    \item \textit{IVDM3Seg}: The Automatic Intervertebral Disc Localization and Segmentation from 3D Multi-modality MR Images (IVDM3Seg) is a collection of 16 manually annotated 3D multi-modal MR scans of the lower spine. Domain adaptation techniques were validated on this dataset~\citep{10.1007/978-3-030-32245-8_37,10.1007/978-3-030-59710-8_48}. The test set is extremely small ($N=4$), and it is a single-class segmentation task. 
    
    \item \textit{MM-WHS}: The Multi-Modality Whole Heart Segmentation (MM-WHS) Challenge 2017 dataset \citep{MM-WHS} is a collection of MRI and CT volumes for cardiac segmentation. Specifically, the training data consist of 20 MRI and 20 unpaired CT volumes with ground truth masks. This dataset has been used to benchmark most multi-classes cross-modality domain adaptation techniques \citep{dou2018unsupervised,10.1007/978-3-030-32245-8_74,CUI2021104726,ijcai2020-455}. While the task is challenging, the very limited size of the test set ($N=4$) strongly reduced the statistical power of comparisons.
    
    \item \textit{CHAOS}: The Combined (CT-MR) Healthy Abdominal Organ Segmentation (CHAOS) dataset \citep{CHAOS} corresponds to 20 MR volumes and 30 unpaired CT volumes. Cross-modality domain adaptation techniques have been tested on this dataset \citep{8988158,10.1007/978-3-030-59713-9_34}. 4 and 6 scans are respectively used as test sets for the MR and CT domains. Consequently, the test sets are particularly small.
    
    \item \textit{BraTS}: The Brain Tumor Segmentation (BraTS) benchmark \citep{menze2014multimodal,bakas2017advancing,BRATS} is a popular dataset for the segmentation of brain tumour sub-regions. While images were collected from a large number of medical institutions with different imaging parameters, the origin of the imaging data is not specified for each case. Instead, unsupervised pathology domain adaptation (high to low grades) has been tested on this dataset \citep{10.1007/978-3-030-33391-1_4}, which is a different problem than ours. Alternatively, BraTS has been used for cross-modality domain adaptation \citep{ijcai2020-455}. However, the problem is artificially generated by removing image modalities and has limited clinical relevance.
    
\end{itemize}

In conclusion, test sets used to assess segmentation methods for unsupervised domain adaptation are either private, small or single-class.

\section{Challenge description}\label{sec:challenge_description}

\subsection{Overview}
The goal of the crossMoDA challenge was to benchmark new and existing unsupervised cross-modality domain adaptation techniques for medical image segmentation. The proposed segmentation task focused on two key brain structures involved in the follow-up and treatment planning of vestibular schwannoma (VS): the tumour and the cochleas. Participants were invited to submit algorithms designed for inference on high-resolution T2 (\Ttwo) scans. Participants had access to a training set of high-resolution T2 scans without their manual annotations. Conversely, manual annotations were provided for an unpaired training set of contrast-enhanced T1 (\Tone) scans. Consequently, the participants had to perform unsupervised cross-modality domain adaptation from \Tone (source) to \Ttwo (target) scans.

\subsection{Data description}
\subsubsection{Data overview}
The dataset for the crossMoDA challenge is an extension of the publicly available Vestibular-Schwannoma-SEG collection released on The Cancer Imaging Archive (TCIA) \citep{ScientificDataShapey2021,clarkEtAl2013TCIA}.
To ensure that no data in the test set was accessible to the participants, no publicly available scan was included in the test set. The open Vestibular-Schwannoma-SEG dataset was used for training and validation, while an extension was kept private and used as the test set.  

The complete crossMoDA dataset (training, validation and testing) contained a set of MR images collected on 379 consecutive patients (Male:Female 166:214; median age: 56 yr, range: 24 yr to 84 yr) with a single sporadic VS treated with Gamma Knife stereotactic radiosurgery (GK SRS) between 2012 and 2021 at a single institution. For each patient, contrast-enhanced T1-weighted (\Tone) and high-resolution T2-weighted (\Ttwo) scans were acquired in a single MRI session prior to and typically on the day of the radiosurgery. 75 patients had previously undergone surgery. 
Data were obtained from the Queen Square Radiosurgery Centre (Gamma Knife). All contributions to this study were based on approval by the NHS Health Research Authority and Research Ethics Committee (18/LO/0532) and were conducted in accordance with the 1964 Declaration of Helsinki. 

The scans acquired between October 2012 and December 2017 correspond to the publicly available Vestibular-Schwannoma-SEG dataset on TCIA (242 patients).
The Vestibular-Schwannoma-SEG dataset was randomly split into three sets: the source training set (105 annotated \Tone scans), the target training set (105 non-annotated \Ttwo scans) and the target validation set (32 non-annotated \Ttwo scans).

The scans acquired between January 2018 and March 2021 were used to make up the test set (137 non-annotated \Ttwo scans). The test set remained private to the challenge participants and accessible only to the challenge organisers, even during the evaluation phase.

As shown in Table~\ref{tab:DataCharacteristics}, the target training, validation, and test sets have a similar distribution of features (age, gender and operative status of patients; slice thickness and in-plane resolution of \Ttwo).

\begin{table*}[tb]
	\centering
	\caption{Summary of data characteristics of the crossMoDA sets
	}\label{tab:DataCharacteristics}
	\resizebox{\textwidth}{!}{
	\begin{tabular}{l *{9}{c}}
		\toprule
		\multirow{2}{*}{} & \multicolumn{3}{c}{\bf Training} & \multicolumn{2}{c}{\bf Validation } & \multicolumn{2}{c}{\bf Test }\\ 
		
       \cmidrule(lr){2-4} \cmidrule(lr){5-6} \cmidrule(lr){7-8}
       & Source & \multicolumn{2}{c}{Target}  & \multicolumn{2}{c}{Target} & \multicolumn{2}{c}{Target}  \\
       \cmidrule(lr){2-2} \cmidrule(lr){3-4} \cmidrule(lr){5-6}
       \cmidrule(lr){7-8}
       
		Sequence & MP-RAGE \Tone{} & CISS \Ttwo & TSE \Ttwo &  CISS \Ttwo & TSE \Ttwo & CISS \Ttwo & TSE \Ttwo \\ 
		
		\rowcolor{LightGray} 
        Number of scans  & 105 & 83  & 22 & 28 & 4 & 132 & 5  \\
        Number of patients  & 105 & 83 & 22 & 28 & 4 & 132 & 5  \\

        \rowcolor{LightGray}
        Available annotations & VS + Cochleas & $\times$ & $\times$  & $\times$ & $\times$ & $\times$  & $\times$
        \\
        & & & & & &  $448\times448$ ($96\%$) & $384\times384$ ($60\%$) \\
        \multirow{-2}{*}{In-plane matrix} & \multirow{-2}{*}{$512\times512$} & \multirow{-2}{*}{$448\times448$} & \multirow{-2}{*}{$384\times384$} & \multirow{-2}{*}{$448\times448$} & \multirow{-2}{*}{$384\times384$} & $512\times512$ ($4\%$) & $512\times512$ ($40\%$) \\
        
        \rowcolor{LightGray} 
        Average axial slice number & $123 \pm 11$ & $80 \pm 1$ & $39 \pm 4$ & $80 \pm 0$ & $40 \pm 0$ & $80 \pm 4$ & $35 \pm 8$ \\

        In-plane resolution in mm & $0.41\times0.41$ & $0.46\times0.46$ & $0.55\times0.55$ & $0.46\times0.46$ & $0.55\times0.55$ & $0.46\times0.46$ & $0.55\times0.55$ \\
        
        \rowcolor{LightGray} 
         & $1.0$ ($7\%$)   & $1.0$ ($10\%$)  &  & $1.0$ ($7\%$) & & $1.0$ ($7\%$) & \\
        \rowcolor{LightGray} 
        \multirow{-2}{*}{Slice thickness in mm}& $1.5$ ($93\%$) & $1.5$ ($90\%$) & \multirow{-2}{*}{$1.5$}  & $1.5$ ($93\%$) & \multirow{-2}{*}{$1.5$} &  $1.5$ ($93\%$) &  \multirow{-2}{*}{$1.5$} \\

        Male:Female & $44\%:56\%$ & \multicolumn{2}{c}{$36\%:64\%$} & \multicolumn{2}{c}{$34\%:66\%$} & \multicolumn{2}{c}{$51\%:49\%$}\\
        \rowcolor{LightGray} 
        Post-operative cases & $26\%$ & \multicolumn{2}{c}{$20\%$} & \multicolumn{2}{c}{$16\%$} & \multicolumn{2}{c}{$15\%$} \\
        
        Age in years - Median [Q1-Q3] & 54 [47-66] & \multicolumn{2}{c}{56 [44-64]} & \multicolumn{2}{c}{58 [51-66]}& \multicolumn{2}{c}{56 [45-66]} \\
		\bottomrule
	\end{tabular}
	}
\end{table*}

\subsubsection{Image acquisition}
All images were obtained on a 32-channel Siemens Avanto 1.5T scanner using a Siemens single-channel head coil. Contrast-enhanced T1-weighted imaging was performed with an MP-RAGE sequence (in-plane resolution=0.47$\times$0.47mm, matrix size=512$\times$512, TR=1900ms, TE=2.97ms) and slice thickness of 1.0-1.5mm.  
High-resolution T2-weighted imaging was performed with either a Constructive Interference Steady State (CISS) sequence (in-plane resolution=0.47$\times$0.47mm, matrix size=448$\times$448, TR=9.4ms, TE=4.23ms) or a Turbo Spin Echo (TSE) sequence (in-plane resolution=0.55$\times$0.55mm, matrix size=384$\times$384, TR=750ms, TE=121ms) and slice thickness of 1.0-1.5mm. The details of the dataset are given in Table~\ref{tab:DataCharacteristics}, and sample cases from the source and target sets are illustrated in Fig~\ref{fig:data_set_pres}.

\subsubsection{Annotation protocol}
All imaging datasets were manually segmented following the same annotation protocol. 

The tumour volume (VS) was manually segmented by the treating neurosurgeon and physicist using both the \Tone and \Ttwo images. All VS segmentations were performed using the Leksell GammaPlan software that employs an in-plane semi-automated segmentation method. Using this software, delineation was performed on sequential 2D axial slices to produce 3D models for each structure. 

The adjacent cochlea (hearing organ) is the main organ at risk during VS radiosurgery. In the crossMoDA dataset, patients have a single sporadic VS. Consequently, only one cochlea per patient - the closest one to the tumour - was initially segmented by the treating neurosurgeon and physicist. Preliminary results using a fully-supervised approach \citep{isensee2021nnu} showed that considering the remaining cochlea as part of the background leads to poor performance for cochlea segmentation. Given that tackling this challenging issue is beyond the scope of the challenge, both cochleas were manually segmented by radiology fellows with over 3 years of clinical experience in general radiology using the ITK-SNAP software \citep{yushkevich2019user}. \Ttwo images were used as reference for cochlea segmentation. The basal turn with osseous spiral lamina was included in the annotation of every cochlea to keep manual labels consistent. In addition, modiolus, a small low-intensity area (on \Ttwo) within the centre of the cochlea, was included in the segmentation as well.

\subsubsection{Data curation}
The data was fully de-identified by removing all health information identifiers and defaced \citep{Milchenko2013}. Details can be found in \cite{ScientificDataShapey2021}. Since the data was acquired consistently (similar voxel spacing, same scanner), no further image pre-processing was employed. Planar contour lines (DICOM RT-objects) of the VS were converted into label maps using SlicerRT~\citep{Pinter2012b}.

Images and segmentation masks were distributed as compressed NIfTI files (.nii.gz). The training and validation data was made available on zenodo\footnote{\url{https://zenodo.org/record/4662239}}. As we expect this dataset to be used for other purposes in addition to cross-modality domain adaptation, the data was released under a permissive copyright-license (CC-BY-4.0), allowing for data to be shared, distributed and improved upon. 

\subsection{Challenge setup}
The validation phase was hosted on Grand Challenge\footnote{\url{https://grand-challenge.org/}}, a well-established challenge platform, allowing for automated validation leaderboard management. Participant submissions are automatically evaluated using the \verb+evalutils+\footnote{\url{https://evalutils.readthedocs.io/en/latest/}} and \verb+MedPy+\footnote{\url{https://loli.github.io/medpy/}} Python packages. To mitigate the risk that participants select their model hyper-parameters in a supervised manner, i.e. by computing the prediction accuracy, only one submission per day was allowed on the validation leaderboard. The validation phase was held between the 5th of May 2021 and the 15th of August 2021.

Following the best practice guidelines for challenge organisation \citep{MAIERHEIN2020101796}, the test set remained private to reduce the risk of cheating. Participants had to containerise their methods with Docker following guidelines\footnote{\url{https://crossmoda.grand-challenge.org/submission/}} and submit their Docker container for evaluation on the test set. Only one submission was allowed. Docker containers were run on a Ubuntu (20.04) desktop with 64GB RAM, an Intel Xeon CPU E5-1650 v3 and an NVIDIA TITAN X GPU with 12 GB memory. To test the quality of the predictions performed on the local machine, predictions on the validation set were computed and compared with the ones obtained using participants' machines. \textit{In fine}, all participant containers passed the quality control test.

\section{Metrics and evaluation}\label{sec:evaluation}
The choice of the metrics used to assess the performance of the participants' algorithm and the ranking strategy are keys for adequate interpretation and reproducibility of results \citep{Maier-Hein2018}. In this section, we follow the BIAS best practice recommendations for assessing challenges \citep{MAIERHEIN2020101796}

\subsection{Choice of the metrics}
The algorithms' main property to be optimised is the accuracy of the predictions. As relying on a single metric for the assessment of segmentations leads to less robust rankings, two metrics were chosen: the Dice similarity coefficient (DSC) and the Average symmetric surface distance (ASSD). DSC and ASSD have frequently been used in previous challenges \citep{CHAOS,antonelli2021medical} because of their simplicity, their rank stability and their ability to assess segmentation accuracy.

Let $S_k$ be the predicted binary segmentation mask of the region $k$ where $k\in\{ \text{VS}, \text{Cochleas} \}$. Let $G_k$ be the manual segmentation of the region $k$. The Dice Score coefficient quantifies the similarity of two masks $S_k$ and $G_k$ by normalising the size of their intersection over the average of their sizes:
\begin{equation}
    \DSC(S_k,G_k) = \frac{2\sum_{i}S_{k,i}G_{k,i}}{\sum_{i}S_{k,i} + \sum_{i}G_{k,i}}
\end{equation}
Let $B_{S_k}$ and $B_{G_k}$ be the boundaries of the segmentation mask $S_k$ and the manual segmentation $G_k$.
The average symmetric surface distance (ASSD) is the average of all the Euclidean
distances (in $\text{mm}$) from points on the boundary $B_{S_k}$ to the boundary $B_{G_k}$ and from points on the boundary of $B_{G_k}$ to the boundary $B_{S_k}$:

\begin{equation}
    \ASSD(S_k,G_k) = \frac{\sum_{s_i\in B_{S_k}}d(s_i, B_{G_k}) +  \sum_{s_i\in B_{G_k}}d(s_i, B_{S_k})}{|B_{S_k}|+|B_{G_k}|}
\end{equation}
where $d$ is the Euclidean distance.

Note that if predictions only contain background, i.e. for all voxels $i$, $S_{k,i}=0$, then the ASSD is set as the maximal distance between voxels in the test set ($350\text{mm}$).

\subsection{Ranking scheme}
We used a standard ranking scheme that has previously been employed in other challenges with satisfactory results, such as the BraTS \citep{BRATS} and the ISLES \citep{MAIER2017250} challenges. Participating teams are ranked for each testing case, for each evaluated region (i.e., VS and cochleas), and for each measure (i.e., DSC and ASSD). The lowest rank of tied values is used for ties (equal scores for different teams). Rank scores are then calculated by firstly averaging across all these individual rankings for each case (i.e., cumulative rank) and then averaging these cumulative ranks across all patients for each participating team. Finally, teams are ranked based on their rank score. This ranking scheme was defined, released prior to the start of the challenge, and available on the dedicated Grand Challenge page\footnote{\url{https://crossmoda.grand-challenge.org/}} and the crossMoDA website\footnote{\url{https://crossmoda-challenge.ml/}}.

To analyse the stability of the ranking scheme, we employed the bootstrapping method detailed in \cite{Wiesenfarth2021}. One bootstrap sample consists of N=137 test cases randomly drawn with replacement from the test set of size N=137. On average, $63\%$ of distinct cases are retained in a bootstrap sample. A total of 1,000 of these bootstrap samples were drawn, and the proposed ranking scheme was applied to each bootstrap sample. The original ranking computed on the test set was then pairwise compared to the rankings based on the individual bootstrap samples. The correlation between these pairs of rankings was computed using Kendall's $\tau$, which provides values between $-1$ (for reverse ranking order) and $1$ (for identical ranking order).

\section{Participating methods}\label{sec:participating_methods}
A total of 341 teams registered to the challenge, allowing them to download the data. 55 teams from 16 different countries submitted predictions to the validation leaderboard. Among them, 16 teams from 9 different countries submitted their containerised algorithm for the evaluation phase. 

In this section, we provide a summary of the methods used by these 16 teams. Each method is assigned a unique colour code used in the tables and figures. Brief comparisons of the proposed techniques in terms of methodology and implementation details (training strategy, pre-, post-processing, data augmentation) are presented in Table~\ref{tab:comparision_teams}. 

To bridge the domain gap between the source and target images, proposed techniques can be categorised into three groups that use:
\begin{enumerate}
    \item Image-to-image translation approaches such as CycleGAN and its extensions to transform \Tone scans into pseudo-\Ttwo scans  (\circle{Samoyed-c}\circle{PKU-BIALAB-c}\circle{jwc-rad-c}\circle{MIP-c}\circle{PremiLab-c}\circle{Epione-Liryc-c}\circle{MedICL-c}\circle{DBMI-pitt-c}\circle{smriti161096-c}\circle{gabybaldeon-c}) or \Ttwo scans into pseudo-\Tone scans (\circle{Hi-Lib-c}). Then, one or multiple segmentation networks are trained on the pseudo-scans using the manual annotations and used to perform image segmentation on the target images.
    \item MIND cross-modal features \citep{HEINRICH20121423} to translate the target and source images in a modality-agnostic feature space. These features are either used to propagate labels using image registration (\circle{IMI-c}) or to train an ensemble of segmentation networks (\circle{GapMIND-c}).
    \item discrepancy measurements either based on discriminative losses (\circle{skjp-c}\circle{IRA-c}) or minimal-entropy correlation (\circle{SEU-Chen-c}) to align features extracted from the source and target images.
    
\end{enumerate}

\begin{sidewaystable*}
	\centering
	\caption{Metrics values and corresponding scores of submission. Median and interquartile values are presented. The best results are given in bold. Arrows indicate favourable direction of each metric.
	}\label{tab:comparision_teams}
	\resizebox{\textwidth}{!}{
	\begin{tabular}{l  *{12}{c}}
		\toprule
		\multirow{2}{*}{} & \multicolumn{6}{c}{\bf Methodology } & \multicolumn{3}{c}{\bf Training Strategy (Segmentation network) } & \multicolumn{3}{c}{\bf Inference Strategy }\\ 
		
       \cmidrule(lr){2-7} \cmidrule(lr){8-10} \cmidrule(lr){11-13}
       & Segmentation  & Feature  & Cross-modal  & Image-to-image & Self-  & & Data & Loss  &  & Pre-processing &  &    \\
       & network & alignment & descriptors & translation & training & \multirow{-2}{*}{Cropping} &  Augmentation & Function(s) & \multirow{-2}{*}{Optimisation} & voxel size (mm) & \multirow{-2}{*}{Ensembling} & \multirow{-2}{*}{Post-processing}   \\
       
		\midrule
         & 3D and 2D &  &  & CycleGAN w. &  &  & & Dice +  & SGD & & $5 \times$ 3D nnU-Net &  \\ 
        \multirow{-2}{*}{\circle{Samoyed-c} Samoyed } & nnU-Net& \multirow{-2}{*}{$\times$} & \multirow{-2}{*}{$\times$} & segme. decoder & \multirow{-2}{*}{$\checkmark$} & \multirow{-2}{*}{Fixed size} & \multirow{-2}{*}{nnU-Net augm.$^1$}  & C-E & Batch:2 & \multirow{-2}{*}{nnU-Net pre$^2$} & $5 \times$ 2D nnU-Net & \multirow{-2}{*}{VS: Largest CC}\\ 

        \rowcolor{LightGray}
        &  &  &  &  &   &  &  & Dice +  & SGD & &  & \\ 
        \rowcolor{LightGray}
        \multirow{-2}{*}{\circle{PKU-BIALAB-c} PKU\textunderscore BIALAB} & \multirow{-2}{*}{3D nnU-Net} & \multirow{-2}{*}{$\times$} & \multirow{-2}{*}{$\times$} & \multirow{-2}{*}{NiceGAN (2D)} & \multirow{-2}{*}{$\checkmark$} & \multirow{-2}{*}{Fixed size} & \multirow{-2}{*}{nnU-Net augm.$^1$}  & C-E & Batch:4 & \multirow{-2}{*}{nnU-Net pre$^2$} & \multirow{-2}{*}{$5 \times$ 3D nnU-Net} & \multirow{-2}{*}{VS: Largest CC}\\ 
    
     &  &  &  &  &   &  & nnU-Net augm.$^1$ & Dice +  & SGD & nnU-Net pre$^2$ &  &  \\ 
    \multirow{-2}{*}{\circle{jwc-rad-c} jwc-rad} & \multirow{-2}{*}{3D nnU-Net} & \multirow{-2}{*}{$\times$} & \multirow{-2}{*}{$\times$} & \multirow{-2}{*}{CUT (2D)} & \multirow{-2}{*}{$\times$} & \multirow{-2}{*}{Fixed size} & VS: intensity augm.   & C-E & Batch: 2 & $0.6\times 0.6 \times 1.0$ & \multirow{-2}{*}{$5 \times$ 3D nnU-Net} &   \multirow{-2}{*}{VS: Largest CC} \\ 
    
    \rowcolor{LightGray}
     & 2.5D U-Net w. & & & CycleGAN (2D+3D) & & Manual ROI+ & Int. shift, contrast, & Dice +  & Adam & $[0, 1]$ norm.  & $3 \times$ 2.5D U-Net & VS: largest CC\\
    \rowcolor{LightGray}
    \multirow{-2}{*}{\circle{MIP-c} MIP} & Attention & \multirow{-2}{*}{$\times$} & \multirow{-2}{*}{$\times$} & CUT (3D) & \multirow{-2}{*}{$\times$} & rigid registration & affine def. & C-E & Batch: 1 & $0.46\times 0.46 \times 1.5$  & Fusing: clean-lab  & Coch.: 2 largest CC \\ 
    
    &  & Content- & & Content-Style & & & Affine+Elastic & Dice + & Adabelief & $[-1,-1]$ norm. \\
    \multirow{-2}{*}{\circle{PremiLab-c} PremiLab} & \multirow{-2}{*}{DAR-U-Net} & Style & \multirow{-2}{*}{$\times$} &   GAN (3D) &  \multirow{-2}{*}{$\times$} & \multirow{-2}{*}{$\times$} & deformation & Focal & Batch: 2 & $0.41 \times 0.41 \times 1.5$ & \multirow{-2}{*}{$\times$} & \multirow{-2}{*}{VS: Largest CC} \\ 
    
    \rowcolor{LightGray}
    & & & & CycleGAN w. & & MNI registration & Flipping, rotation, & & Adam & $[0,1]$ norm. & & K-means $\&$   \\
    \rowcolor{LightGray}
    \multirow{-2}{*}{\circle{Epione-Liryc-c} Epione-Liryc} &\multirow{-2}{*}{3D U-Net} & \multirow{-2}{*}{$\times$} & \multirow{-2}{*}{$\times$} & Pair-Loss  & \multirow{-2}{*}{$\times$} & label-based ROI & Int. Noise & \multirow{-2}{*}{Dice}& Batch: 1 & $0.5 \times 0.5 \times 0.5$& \multirow{-2}{*}{$\times$} & mean-shift + CRF \\ 
    
    & 2.5D U-Net & & & & & MNI registration & Affine+Elastic & & Adam & $[0,1]$ norm. & $2\times$  2.5D U-Net & \\
    
    \multirow{-2}{*}{\circle{MedICL-c} MedICL} & 3D CNN & \multirow{-2}{*}{$\times$} & \multirow{-2}{*}{$\times$} & \multirow{-2}{*}{CycleGAN (2D)} & \multirow{-2}{*}{$\times$} & label-based ROI & Deformation + IT$^3$ & \multirow{-2}{*}{Dice} & Batch: 2 & $0.38\times0.45\times1.5$ & $2\times $3D CNN & \multirow{-2}{*}{VS: Largest CC} \\ 
    
    \rowcolor{LightGray}
    & 3D U-Net w. & & & & & & Int. shift, resizing, & Attention & Adam & [0,4000] norm & & VS: Largest CC\\
    \rowcolor{LightGray}
    \multirow{-2}{*}{\circle{DBMI-pitt-c} DBMI\textunderscore pitt} & attention &\multirow{-2}{*}{$\times$} &\multirow{-2}{*}{$\times$} & \multirow{-2}{*}{CUT (3D)} &\multirow{-2}{*}{$\times$} & \multirow{-2}{*}{$\times$}   & affine def. & Dice & Batch:4 & $1.0 \times 1.0 \times 1.0$ & \multirow{-2}{*}{$\times$} & Hole filling\\ 
    
    &  & & & CycleGAN (2D) &  & & & & Adam & $[-1,1]$ norm & & Small CC \\
    \multirow{-2}{*}{\circle{Hi-Lib-c} Hi-Lib} & \multirow{-2}{*}{2.5D U-Net} & \multirow{-2}{*}{$\times$} & \multirow{-2}{*}{$\times$} & CUT (2D) & \multirow{-2}{*}{$\times$} & \multirow{-2}{*}{Label-based ROI} & \multirow{-2}{*}{Flipping} & \multirow{-2}{*}{Dice} & Batch:4 & (X-Y) $0.41 \times 0.41$  & \multirow{-2}{*}{$\times$} & removal + CRF\\ 
    
    \rowcolor{LightGray}
    & & & & & & & & Dice + & Adam & Resizing (X-Y): & &\\
    \rowcolor{LightGray}
    \multirow{-2}{*}{\circle{smriti161096-c} smriti161096} & \multirow{-2}{*}{2D nnU-Net} & \multirow{-2}{*}{$\times$} & \multirow{-2}{*}{$\times$} & \multirow{-2}{*}{CycleGAN (2D)} & \multirow{-2}{*}{$\checkmark$} & \multirow{-2}{*}{Label-based ROI} &   \multirow{-2}{*}{nnU-Net augm.$^1$} & C-E & Batch:51 & $192\times224$  & \multirow{-2}{*}{$5\times$ 2D nnU-Net} &  \multirow{-2}{*}{VS: Largest CC}  \\

    & & Registration-  & & & & & & Dice + & Adam \\
    \multirow{-2}{*}{\circle{IMI-c} IMI} & \multirow{-2}{*}{3D nnU-Net} & based & \multirow{-2}{*}{MIND} & \multirow{-2}{*}{$\times$} & \multirow{-2}{*}{$\times$} & \multirow{-2}{*}{Fixed size} & \multirow{-2}{*}{nnU-Net augm.$^1$} & C-E & Batch:1 & \multirow{-2}{*}{nnU-Net pre$^2$} & \multirow{-2}{*}{$5\times$ 3D nnU-Net} & \multirow{-2}{*}{CRF}\\ 
    
    \rowcolor{LightGray}
    & 3D DeepLab w. & & & & &  & Int. noise, & Weighted & Adam & z-score norm & &\\
 
    \rowcolor{LightGray}
    \multirow{-2}{*}{\circle{GapMIND-c} GapMIND} & MobileNetV2 & \multirow{-2}{*}{$\times$} & \multirow{-2}{*}{MIND} &  \multirow{-2}{*}{$\times$} & \multirow{-2}{*}{$\times$} & \multirow{-2}{*}{Half size (X-Y)} & affine def. & C-E & Batch:1 & (X-Y) $0.5\times0.5$ & \multirow{-2}{*}{$\times$} & \multirow{-2}{*}{$\times$} \\ 
    
    & & Adversarial loss& & & & & Flipping, affine+ & Dice + & Adam & $[0,1]$ norm. \\
    \multirow{-2}{*}{\circle{gabybaldeon-c} gabybaldeon} & \multirow{-2}{*}{2D U-Net}  & on outputs & \multirow{-2}{*}{$\times$} & \multirow{-2}{*}{CycleGAN (2D)}  & \multirow{-2}{*}{$\times$} & \multirow{-2}{*}{$\times$} & elastic def. & C-E & Batch: 1 & $0.46 \times 0.46 \times 1.5$ & \multirow{-2}{*}{$\times$} & \multirow{-2}{*}{$\times$} \\ 
    
    \rowcolor{LightGray}
    & & & & & &  & & Weighted & Nesterov & & $2 \times$ 3D nnU-Net & \\
    \rowcolor{LightGray}
    \multirow{-2}{*}{\circle{SEU-Chen-c} SEU\textunderscore Chen} & \multirow{-2}{*}{3D nnU-Net} & \multirow{-2}{*}{Entropy} & \multirow{-2}{*}{$\times$} & \multirow{-2}{*}{$\times$} & \multirow{-2}{*}{$\times$} & \multirow{-2}{*}{$\times$} & \multirow{-2}{*}{nnU-Net augm.$^1$} & C-E & Batch: 2 & \multirow{-2}{*}{nnU-Net pre$^2$} & (source + adapted) & \multirow{-2}{*}{$\times$} \\ 
    
    & 3D E-Net & Gradient  & & & & & & Dice + & Adam & z-score norm \\
    \multirow{-2}{*}{\circle{skjp-c} skjp} & per structure & Reversal Layer & \multirow{-2}{*}{$\times$} & \multirow{-2}{*}{$\times$} & \multirow{-2}{*}{$\times$} & \multirow{-2}{*}{$\times$} & \multirow{-2}{*}{$\times$} & C-E & Batch: 4+16 & $0.5 \times 0.5 \times 0.5$ & \multirow{-2}{*}{$\times$} & \multirow{-2}{*}{$\times$} \\ 
    
    \rowcolor{LightGray}
    &  & Gradient & & & & & Int. Gamma- & Dice + & Adam & $[0,1]$ norm. & & \\
    \rowcolor{LightGray}
    \multirow{-2}{*}{\circle{IRA-c} IRA} & \multirow{-2}{*}{3D U-Net}& Reversal Layer & \multirow{-2}{*}{$\times$} & \multirow{-2}{*}{$\times$} & \multirow{-2}{*}{$\times$} & \multirow{-2}{*}{Fixed size} & transform & C-E & Batch: 2 & $0.5 \times 0.5 \times 0.5$ & \multirow{-2}{*}{$\times$} & \multirow{-2}{*}{$\times$} \\

	\bottomrule
	\end{tabular}
	}
\smallskip
\parbox[t]{\textwidth}{\footnotesize
  $^1$: Rotations, scaling, Gaussian noise, Gaussian blur, brightness, contrast, simulation of low resolution, gamma correction and mirroring.\\
  $^2$: Cropping: cropped the background regions of the images so that the images could fit the brains. Resampling: In-plane with third-order spline, out-of-plane with nearest neighbour Intensity normalisation: z-score normalisation.
  \\
  $^3$ Intensity Augmentation (IT): random apply multi-channel Contrast Limited Adaptive Equalization (mCLAHE) and gamma correction, Gaussian blur, Gaussian noise and image sharpening.
  \\
  def.: deformations; augm.: augmentation; Int.: intensity; C-E: cross-entropy; CC: connected component.
}
\end{sidewaystable*}

Each of the methods is now succinctly described with reference to a corresponding paper whenever available.

\paragraph{\textbf{\circle{Samoyed-c}  Samoyed (1st place, Shin et al.)}}
The proposed model is based on target-aware domain translation and self-training \citep{shin2021self}. Labelled \Tone scans are first converted to pseudo-\Ttwo scans using a modified version of CycleGAN \citep{zhu2017unpaired}, where an additional decoder is attached to the shared encoder to perform vestibular schwannoma and cochleas segmentation simultaneously with domain conversion, thereby preserving the shape of vestibular schwannoma and cochleas in the generated pseudo-\Ttwo scans. Next, self-training is employed, which consists of 1) training segmentation with labelled pseudo-\Ttwo scans, 2) inferring pseudo-labels on unlabelled real \Ttwo scans by using the trained model, and 3) retraining segmentation with the combined data of labelled pseudo-\Ttwo scans and pseudo-labelled real \Ttwo scans. For self-training, nnU-Net \citep{isensee2021nnu} is used as the backbone segmentation model. 2D and 3D models are ensembled, and all-but-largest-component-suppression is applied to vestibular schwannoma.

\paragraph{\textbf{\circle{PKU-BIALAB-c}  PKU\textunderscore BIALAB (2nd place, Dong et al.)}} \cite{PAST} proposed an unsupervised cross-modality domain adaptation approach based on pixel alignment and self-training (PAST). During training, pixel alignment is applied to transfer \Tone scans to \Ttwo modality to alleviate the domain shift. The synthesised \Ttwo scans are then used to train a segmentation model with supervised learning. To fit the distribution of \Ttwo scans, self-training \citep{DAST} is applied to adapt the decision boundary of the segmentation network. The model in the pixel alignment stage relies on NiceGAN \citep{chen2020reusing} (i.e., an extension method of CycleGAN), which improves the efficiency and effectiveness of training by reusing discriminators for encoding. For 3D segmentation, the nnU-Net~\citep{isensee2021nnu} framework is used with the default 3D full resolution configuration.

\paragraph{\textbf{\circle{jwc-rad-c} jwc-rad (3rd place, Choi)}} The proposed method \citep{choi2021using} is based on out-of-the-box deep learning frameworks for unpaired image translation and image segmentation. For domain adaptation, CUT \citep{park2020contrastive}, a model for unpaired image-to-image translation based on patch-wise contrastive learning and adversarial learning, is used. CUT was implemented using the default configurations of the framework except that no resizing or cropping was performed, and the number of epochs with the initial learning rate and the number of epochs with decaying learning rate were both set to 25. For the segmentation task, nnU-Net \citep{isensee2021nnu} is used with the default 3D full resolution configuration of the framework, except that the total number of epochs for training is set to 250. Data augmentation for the segmentation task is performed by generating additional training data with lower tumour signals by reducing the signal intensity of the labelled vestibular schwannomas by $50\%$.

\paragraph{\textbf{\circle{MIP-c} MIP (4th place, Liu et al.)}}
This team proposed to minimise the domain divergence by image-level domain alignment \citep{liu2021crossmodality}. The target domain pseudo-images are synthesised and used to train a segmentation model with source domain labels. Three image translation models are trained, including 2D/3D CycleGANs and a 3D Contrastive Unpaired Translation (CUT) model. The segmentation backbone follows the same architecture proposed in \cite{10.1007/978-3-030-32245-8_96}. To improve the segmentation performance, the segmentation model is fine-tuned using both labelled pseudo-T2 images and unlabelled real T2 images via a semi-supervised learning method called Mean Teacher \citep{NIPS2017_68053af2}. Lastly, the predictions from three models are fused by a noisy label correction method named CLEAN-LAB \citep{10.1613/jair.1.12125}. Specifically, the softmax output from one model is converted to a one-hot encoded mask, which is considered to be a "noisy label" and corrected by the softmax output from another model. For pre-processing, the team manually determined a bounding box around the cochleas as a ROI in an atlas (randomly selected volume) and obtained ROIs in the other volumes by rigid registration. 
For post-processing, the tumour components with centres 15 pixels superior to the centres of cochleas are considered to be false positive and thus removed. Moreover, 3D connected components analysis was utilised to ensure that only two cochleas and one tumour are remained.

\paragraph{\textbf{\circle{PremiLab-c} PremiLab (5th place, Yao et al.)}}
The proposed framework consists of a content-style disentangled GAN for style transfer and a modified 3D version ResU-Net along with two types of attention modules for segmentation (DAR-U-Net). Specifically, content is extracted from both modalities using the same encoder, while style is extracted using modality-specific encoders. A discriminative approach is adopted to align the content representations of the source and target domain. Once the GAN is trained, \Tone-to-\Ttwo images are generated with diverse styles to later train the segmentation network, which can imitate the diversity of \Ttwo domain. Different from the original 2D ResU-Net \citep{diakogiannis2020resunet}, 2.5D structure and group normalisation are employed for computation efficiency of 3D images. Meanwhile, Voxel-wise Attention Module (VAM) and Quartet Attention Module (QAM) are implemented in each level of the decoder and each residual block, respectively. VAM enhances the essential areas of the feature maps in the decoder using the encoder feature, while QAM captures the inter-dimensional dependencies to improve networks with low computation cost. Code available at: \url{https://github.com/Kaiseem/DAR-UNet}.

\paragraph{\textbf{\circle{Epione-Liryc-c} Epione-Liryc (6th place, Ly et al.)}} The team proposed a regularised image-to-image translation approach. First, input images are spatially normalised to MNI space using SPM12\footnote{https://www.fil.ion.ucl.ac.uk/spm/software/spm12/}, allowing for the identification of a global region of interest bounding box using the simple addition of the ground truth labels. The cross-modality domain adaptation is then performed using a CycleGAN model \citep{zhu2017unpaired} to translate the \Tone to pseudo-\Ttwo.
To improve the performance of the CycleGAN, a supervised regularisation technique is added to control the training process, called the Pair-Loss. This loss is calculated as the MSE loss between the pairs of closest 2D-slice images, which are semi-automatically selected using the cross-entropy metric.
The segmentation model is built using the 3D U-Net architecture and trained using both \Tone and pseudo-\Ttwo data. 
At the inference stage, the segmentation output is reverted to the original spatial domain and refined using majority voting between the segmented mask and the k-Means and Mean-Shift derived masks. Finally, Dense Conditional Random Field (CRF) \citep{NIPS2011_beda24c1} is applied to further improve the segmentation result.

\paragraph{\textbf{\circle{MedICL-c} MedICL (7th place, Li et al.)}}
This framework proposed by \cite{li2021unsupervised} consists of two components: Synthesis and segmentation. For the synthesis component, the CycleGAN pipeline is used for unpaired image-to-image translation between \Tone and \Ttwo MRIs. For the segmentation component, the generated \Ttwo MRIs are fed into two 2.5D U-Net models \citep{10.1007/978-3-030-32245-8_96} and two 3D U-Net models \cite{li2021mri}. The 2.5D models contain both 2D and 3D convolutions, as well as an attention module. Residual blocks and deep supervision are used in the 3D CNN models. Furthermore, various data augmentation schemes are applied during training to cope with MRIs from different scanners, including spatial, image appearance, and image quality augmentations. Different parameter settings are used for the two 2.5D CNN models. The difference between the two 3D CNN models is that only one of them had an attention module. Finally, the models are ensembled to obtain the final segmentation result.

\paragraph{\textbf{\circle{DBMI-pitt-c} DBMI\textunderscore pitt (8th place,  Zu et al.)}}
The proposed framework use 3D image-to-image translation to generate pseudo-data used to train a segmentation network. To perform image-to-image translation, authors extended the 2D CUT model \citep{park2020contrastive} to 3D translation.  The translation model consists of a generator $G$, a discriminator $D$ and a feature extractor $F$. The generator $G$ is built upon the 2.5D attention U-Net proposed in \cite{10.1007/978-3-030-32245-8_96}, where two down-sampling layers are removed. For the discriminator $D$, the PatchGAN discriminator is selected \cite{pix2pix2017}. The model structure of $D$ is the 6 layers of Resnet, and when feeding images to the $D$, the images are divided into 16 equal-size patches, which is faster for model feed-forward without sacrificing any performance. The feature extractor $F$ is a simple multi-layer full-connected network as \cite{park2020contrastive}. The segmentation network is a 2.5D \citep{10.1007/978-3-030-32245-8_96}.
To perform image segmentation, the attention 2.5 U-Net \citep{10.1007/978-3-030-32245-8_96} is used as backbone architecture. A detection module is built upon it.
Finally, post-processing is performed using standard morphological operations (hole filling) and the largest component selection for VS. Code is available at: \url{https://github.com/chkzhao/crossMoDA.git}.

\paragraph{\textbf{\circle{Hi-Lib-c} Hi-Lib (9th place,  Wu et al.)}}

The proposed approach is based on the 2.5D attention U-Net~\citep{10.1007/978-3-030-32245-8_96}, where a GAN-based data augmentation strategy is employed to eliminate the instability of unpaired image-to-image translation. Specifically, a source-domain image is sent to the trained CycleGAN~\citep{zhu2017unpaired} and CUT~\citep{park2020contrastive} to obtain two different pseudo-target images, and then they are converted back to source domain-like images so that each source domain image shares the same label with its augmented versions. To pre-process the data, the team calculated the largest bounding box based on the labelled training images and used it to crop all the images. The training process is done in PyMIC, where intensity normalisation, random flipping and cropping are used in training. Each test image is translated into source domain-like by CycleGAN and sent to the trained 2.5D segmentation network. After inference, conditional random fields and removing small-connected regions are used for post-processing. Code available at: \url{https://github.com/JianghaoWu/FPL-UDA.git}.

\paragraph{\textbf{\circle{smriti161096-c} smriti161096 (10th place, Joshi et al.)}}
This approach is based on existing frameworks. The method follows three main steps: pre-processing, image-to-image translation and image segmentation. First, axial slices from MRI are selected using maximum coordinates of bounding boxes across all available segmentation masks and resizing them to a uniform size. Then, pseudo-\Ttwo images are generated using the 2D CycleGAN~\citep{zhu2017unpaired} architecture. Finally, the 3D nnU-Net~\citep{isensee2021nnu} framework is trained using the generated images and their manual annotations. To improve the segmentation, the segmentation model is pre-trained using the \Tone images. Moreover, multiple checkpoints are selected from CycleGAN training to generate images with varying representations of tumours. In addition to this, self-training is employed: pseudo-labels for \Ttwo data are generated with the trained network and then used to further train the network with real images in \Ttwo modality. Lastly, 3D component analysis is applied as a post-processing step to keep the largest connected component for the tumour label.

\paragraph{\textbf{\circle{IMI-c} IMI (11th place, Hansen et al.)}}
The proposed approach is based on robust deformable multi-modal multi-atlas registration to bridge the domain gap between T1 (source) and T2 (target) weighted MRI scans. 
The source and target domain are resampled to isotropic 1~mm resolution and cropped with an ROI of size 64×64×96 voxels within the left and right hemispheres. 30 source training images are randomly selected and automatically registered to a subset of the target training scans both linearly and non-rigidly. Registration is performed using the discrete optimisation framework deeds \cite{heinrich2013mrf} with multi-modal feature descriptors (MIND-SSC \cite{heinrich2013towards}). The propagated source labels are fused using the popular STAPLE algorithm. For fast inference, a nnU-Net model is trained on the noisy labels in the target domain. Based on the predicted segmentations, an automatic centre crop of 48×48×48~mm is chosen (on both hemispheres using the centre-of-mass) with a 0.5~mm resolution. The described process (multi-atlas registration, label propagation and fusion using STAPLE, nnU-Net training) is repeated on the refined crops.

\paragraph{\textbf{\circle{GapMIND-c} GapMIND (12th place, Kruse et al.)}}

The proposed approach uses modality-independent neighbourhood descriptors (MIND) \citep{HEINRICH20121423} to obtain a domain-invariant representation of the source and target data. MIND features describe each voxel with the intensity relations between the surrounding image patches. To perform image segmentation, a Deeplab segmentation pipeline with a MobileNetV2 backbone \citep{Sandler_2018_CVPR} is then trained on the annotated source MIND feature maps obtained from the annotated source images. To improve the performance of the segmentation network, pseudo-labels are generated for the MIND feature maps from the target data and used during training. Specifically, the authors used the approach from the IMI team (\circle{IMI-c}) based on image registration and STAPLE fusion to obtain noisy labels for these target MIND feature maps.

\paragraph{\textbf{\circle{gabybaldeon-c} gabybaldeon (13th place, Calisto et al.)}} This team implemented an image- and feature-level adaptation method \citep{baldeoncalisto2021cmada}. First, images from the source domain are translated to the target domain via a CycleGAN model \citep{zhu2017unpaired}. Then, a 2D U-Net model is trained to segment the target domain images in two stages. In the first stage, the U-Net is trained using the translated source images and their annotations. In the second phase, the feature-level adaptation is achieved through an adversarial learning scheme. The pre-trained U-Net network takes the role of the generator and predicts the segmentations for the target and translated source domain images. Inspired by \cite{li2020domain}, the discriminator takes as input the concatenation of the predicted segmentations, the element-wise multiplication of the predicted segmentation and the original image, and the contour of the predicted segmentation by applying a Sobel operator. This input provides information about the shape, texture, and contour of the segmented region to force the U-Net to be boundary and semantic-aware. 

\paragraph{\textbf{\circle{SEU-Chen-c} SEU\textunderscore Chen (14th place, Xiaofei et al.)}} The proposed approach employs minimal-entropy correlation alignment \citep{morerio2018minimalentropy} to perform domain adaptation. Two segmentation models are trained using the 3D nnU-Net~\citep{isensee2021nnu} framework. Firstly, the annotated source training set is used to train a 3D nnU-net framework. Secondly, another 3D nnU-net framework is trained with domain adaptation. Specifically, DA is performed by minimising the weighted cross-entropy on the source domain and the weighted entropy on the target domain. At the inference stage, VS segmentation is performed using the adapted nnU-Net framework. In contrast, the two models are ensembled for the cochleas task.


\paragraph{\textbf{\circle{skjp-c} skjp (15th place, Kondo)}}
The proposed approach uses Gradient Reversal Layer (GRL)~\citep{ganin2016domain} to perform domain adaption. First, a 3D version of ENet~\citep{paszke2016enet} is trained with 
the annotated source domain dataset. Then, domain adaptation is performed. Specifically, feature maps extracted in the encoder part of the network are fed to GRL. The GRL's output is then used as input of a three fully-connected layers domain classifier. The segmentation network, GRL, and domain classification network are trained with samples from both source and target domains using adversarial learning.
Two separate networks are trained for VS and cochleas segmentation. 

\paragraph{\textbf{\circle{IRA-c} IRA (16th place, Belkov et al.)}}
Gradient Reversal Layer (GRL)~\citep{ganin2016domain} was utilised to perform domain adaption. Two slightly modified 3D U-Net architectures are used to solve the binary segmentation task for cochleas and VS, respectively. These models are trained using \Tone data and adapted using pairs of \Tone and \Ttwo scans.
The adversarial head is used to align the domain features as in \cite{ganin2016domain}. Contrary to the original implementation, the Gradient Reversal Layer is attached to the earlier network blocks based on the layer-wise domain shift visualisation from \cite{zakazov2021anatomy}. 

\begin{table*}[tb]
	\centering
	\caption{Metrics values and corresponding scores of submission. Median and interquartile values are presented. The best results are given in bold. Arrows indicate favourable direction of each metric.
	}\label{tab:Scores}
	\resizebox{\textwidth}{!}{
	\begin{tabular}{l c *{6}{c}}
		\toprule
		\multirow{2}{*}{} & \multicolumn{2}{c}{\bf Challenge Rank } & \multicolumn{2}{c}{\bf Vestibular Schwannoma } & \multicolumn{2}{c}{\bf Cochleas }\\ 
		
       \cmidrule(lr){2-3} \cmidrule(lr){4-5} \cmidrule(lr){6-7}
       & Global Rank $\downarrow$ & Rank Score $\downarrow$ & DSC $(\%)$ $\uparrow$ & ASSD (mm) $\downarrow$ & DSC $(\%)$  $\uparrow$ & ASSD (mm) $\downarrow$ \\
       
		\midrule
		Full-supervision & - & - & 92.5 [89.2 - 94.2]  & 0.20 [0.14 - 0.29] & 87.7 [85.8 - 89.3] & 0.10 [0.09 - 0.13] \\
		\midrule

        \circle{Samoyed-c} Samoyed & \textbf{1} & \textbf{2.7} & 87.0 [81.8 - 90.0] & 0.39 [0.29 - 0.52] & \textbf{85.7 [83.9 - 87.0]} & \textbf{0.13 [0.12 - 0.17]} \\ 
        
        \rowcolor{LightGray}
        \circle{PKU-BIALAB-c} PKU\textunderscore BIALAB & 2 & 3.4 & \textbf{88.4 [84.5 - 91.7]} & \textbf{0.31 [0.23 - 0.4]} & 80.6 [78.0 - 82.5] & 0.18 [0.15 - 0.22] \\ 
        
        \circle{jwc-rad-c} jwc-rad & 3 & 4.2 & 86.2 [81.2 - 90.3] & 0.56 [0.36 - 1.15] & 83.1 [81.0 - 84.6] & 0.17 [0.14 - 0.2] \\ 
        
        \rowcolor{LightGray}
        \circle{MIP-c} MIP & 4 & 4.5 & 83.7 [78.5 - 87.9] & 0.49 [0.38 - 0.67] & 83.2 [80.5 - 85.1] & 0.17 [0.14 - 0.21] \\ 
        
        \circle{PremiLab-c} PremiLab & 5 & 5.2 & 83.3 [77.8 - 88.2] & 0.48 [0.36 - 0.78] & 80.7 [78.3 - 82.5] & 0.19 [0.16 - 0.22] \\ 
        
        \rowcolor{LightGray}
        \circle{Epione-Liryc-c} Epione-Liryc & 6 & 6.0 & 85.3 [77.8 - 88.9] & 0.43 [0.32 - 0.68] & 77.8 [74.3 - 79.6] & 0.24 [0.2 - 0.28] \\ 
        
        \circle{MedICL-c} MedICL & 7 & 6.6 & 84.0 [76.9 - 88.9] & 0.48 [0.35 - 0.62] & 75.2 [71.1 - 78.4] & 0.29 [0.23 - 0.36] \\ 
        
        \rowcolor{LightGray}
        \circle{DBMI-pitt-c} DBMI\textunderscore pitt & 8 & 8.4 & 50.1 [25.7 - 72.0] & 9.23 [1.33 - 11.98] & 81.6 [78.3 - 83.1] & 0.18 [0.15 - 0.23] \\ 
        
        \circle{Hi-Lib-c} Hi-Lib & 9 & 8.8 & 77.3 [58.1 - 85.4] & 1.8 [1.09 - 2.83] & 71.7 [58.6 - 76.8] & 0.28 [0.2 - 0.5] \\ 
        
        \rowcolor{LightGray}
        \circle{smriti161096-c} smriti161096 & 10 & 9.7 & 76.7 [69.4 - 81.2] & 0.69 [0.56 - 0.83] & 51.1 [45.7 - 57.2] & 0.52 [0.43 - 0.63] \\ 
        
        \circle{IMI-c} IMI & 11 & 11.0 & 65.3 [42.8 - 82.6] & 1.26 [0.77 - 1.95] & 45.7 [38.0 - 53.4] & 0.98 [0.59 - 17.24] \\ 
        
        \rowcolor{LightGray}
        \circle{GapMIND-c} GapMIND & 12 & 11.2 & 65.6 [53.6 - 75.5] & 1.64 [1.13 - 4.09] & 53.0 [48.3 - 57.6] & 0.64 [0.58 - 0.79] \\ 
        
        \circle{gabybaldeon-c} gabybaldeon & 13 & 11.9 & 69.5 [55.6 - 79.5] & 4.32 [2.53 - 8.09] & 43.4 [30.8 - 52.5] & 0.9 [0.65 - 1.35] \\ 
        
        \rowcolor{LightGray}
        \circle{SEU-Chen-c} SEU\textunderscore Chen & 14 & 13.2 & 0.0 [0.0 - 18.3] & 33.62 [20.75 - 51.86] & 56.7 [34.8 - 69.6] & 0.62 [0.38 - 15.74] \\ 
        
        \circle{skjp-c} skjp & 15 & 13.9 & 1.7 [0.0 - 38.6] & 12.73 [4.38 - 29.03] & 5.9 [0.0 - 42.5] & 12.54 [1.03 - 26.07] \\ 
        
        \rowcolor{LightGray}
        \circle{IRA-c} IRA & 16 & 14.7 & 0.0 [0.0 - 12.1] & 28.09 [12.87 - 41.92] & 22.6 [3.3 - 34.4] & 14.48 [2.72 - 21.23] \\

		\bottomrule
	\end{tabular}
	}
\end{table*}

\begin{figure*}[p]
\centering
\subfloat[Dice Score Similarity (\%)]{\label{fig:vsdice}{\includegraphics[width=0.48\textwidth]{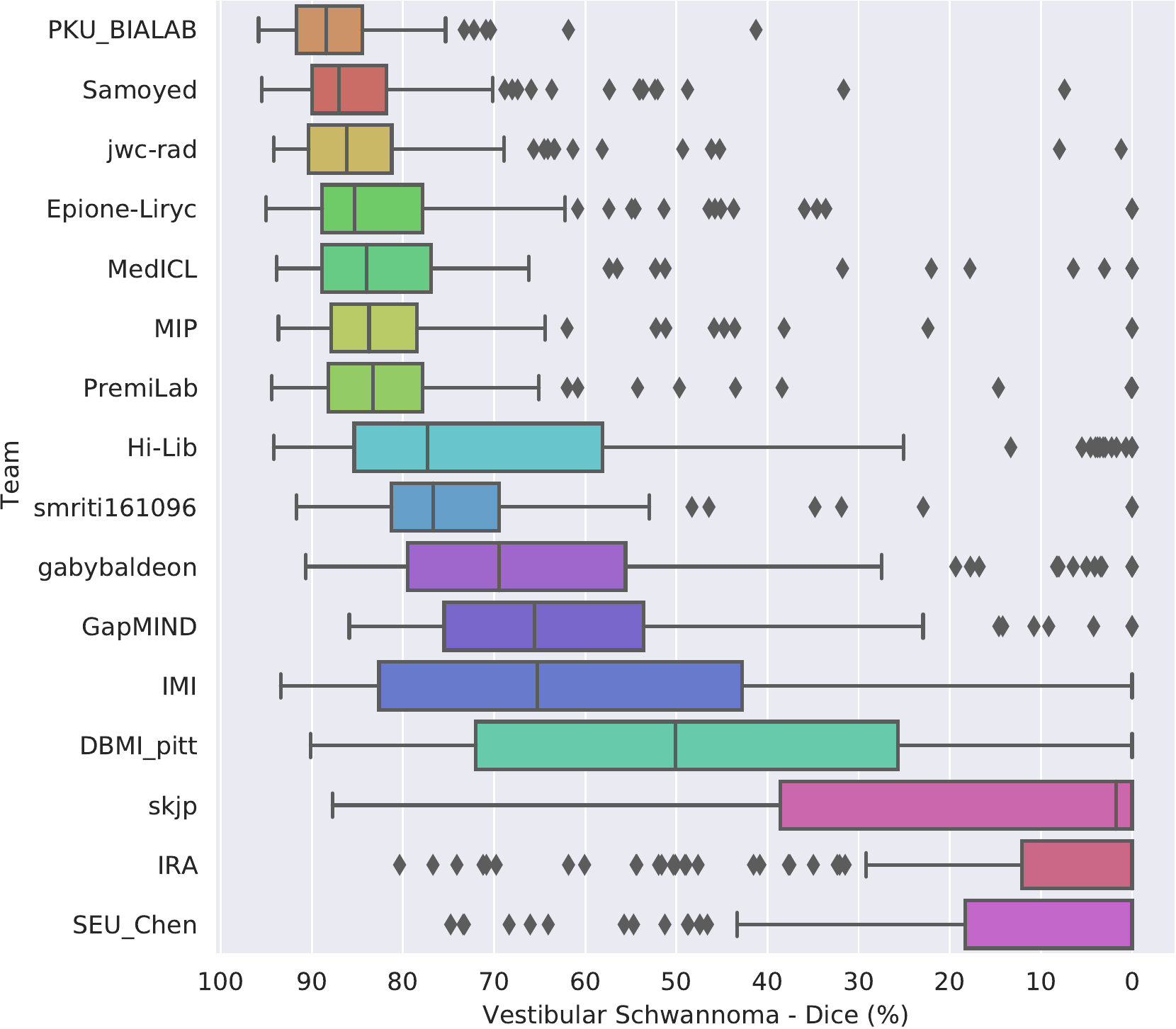}}}\hfill
\subfloat[Average symmetric surface distance (mm)]{\label{fig:vsassd}{\includegraphics[width=0.48\textwidth]{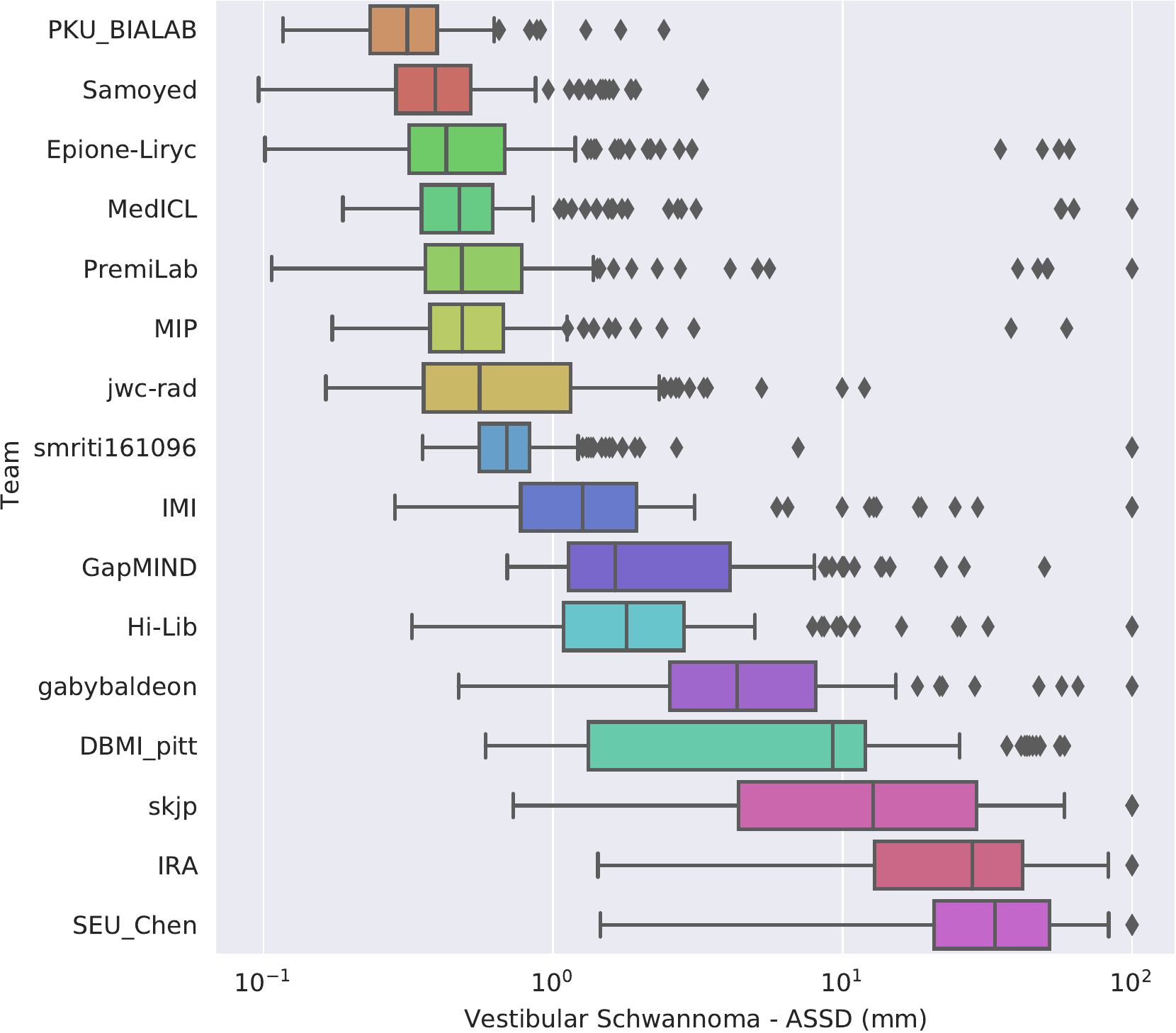}}}
\caption{Box plot of the method's segmentation performance for the vestibular schwannoma  in terms of (a) DSC and (b) ASSD. }
\label{fig:boxvs}

\subfloat[Dice Score Similarity (\%)]{\label{fig:codice}{\includegraphics[width=0.48\textwidth]{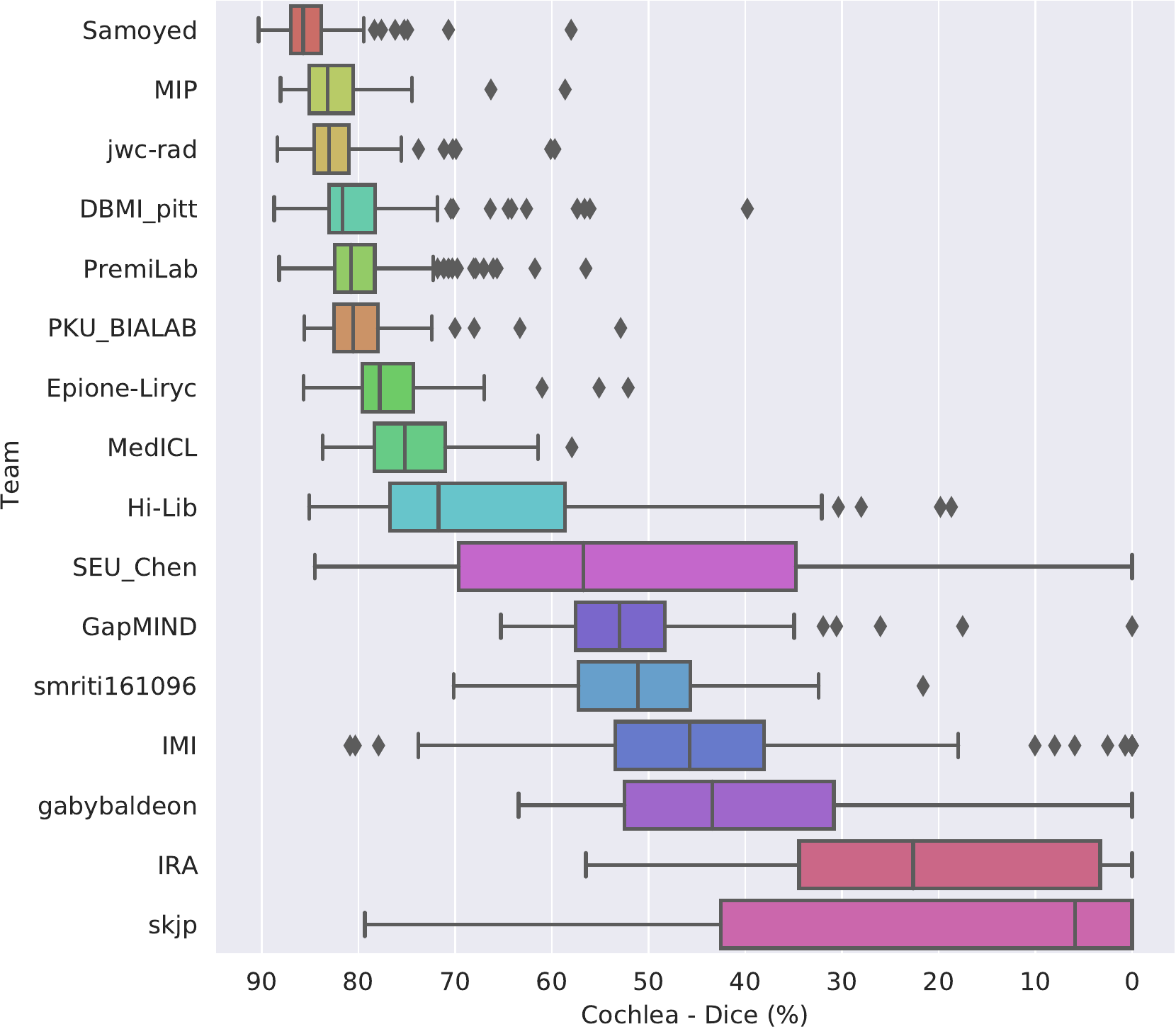}}}\hfill
\subfloat[Average symmetric surface distance  (mm)]{\label{fig:coassd}{\includegraphics[width=0.48\textwidth]{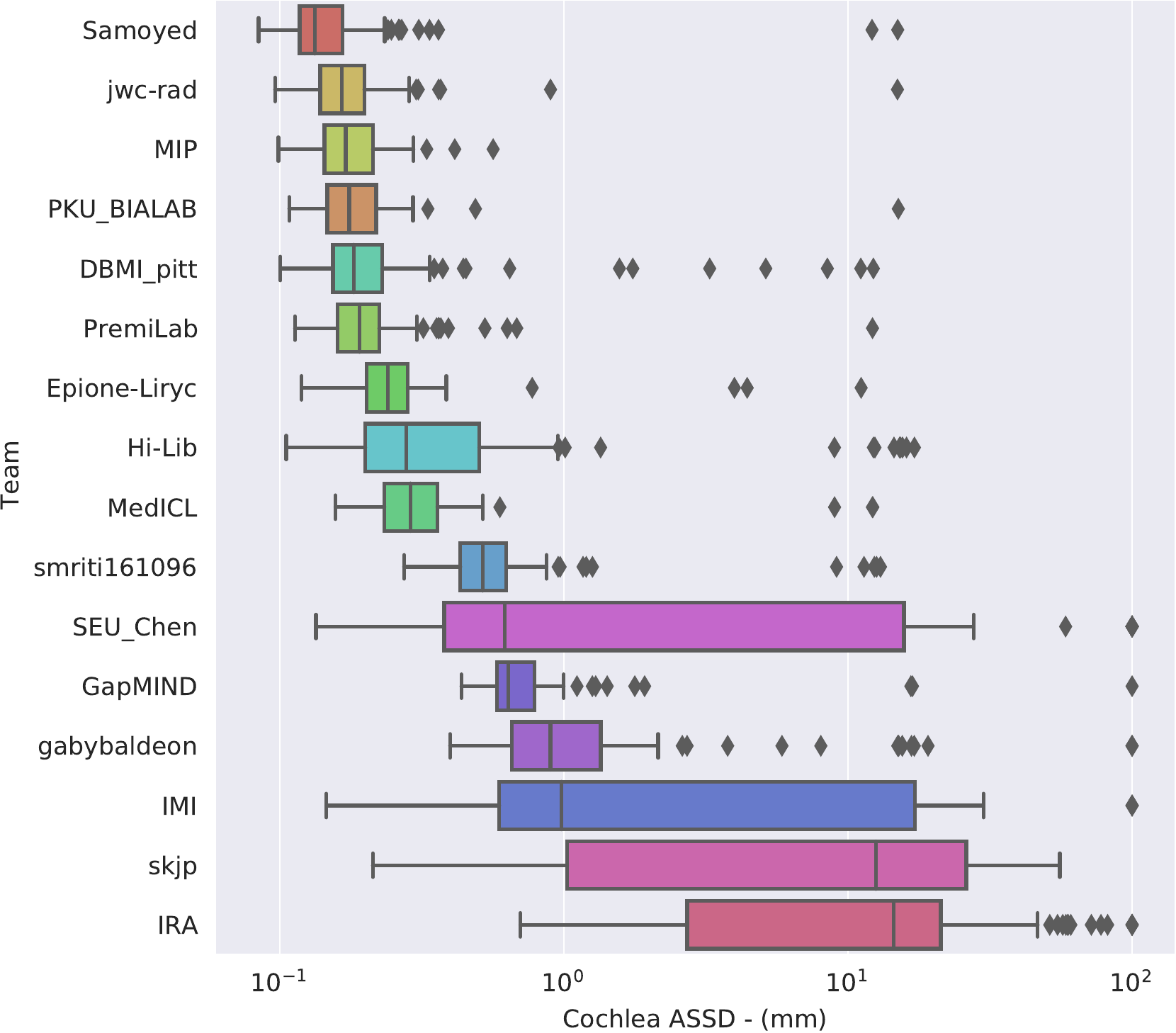}}}
\caption{Box plot of the method's segmentation performance for the cochleas in terms of (a) DSC and (b) ASSD. }

\label{fig:boxco}
\end{figure*}

 \begin{figure*}[h!]
  \centering
  \includegraphics[width=\linewidth]{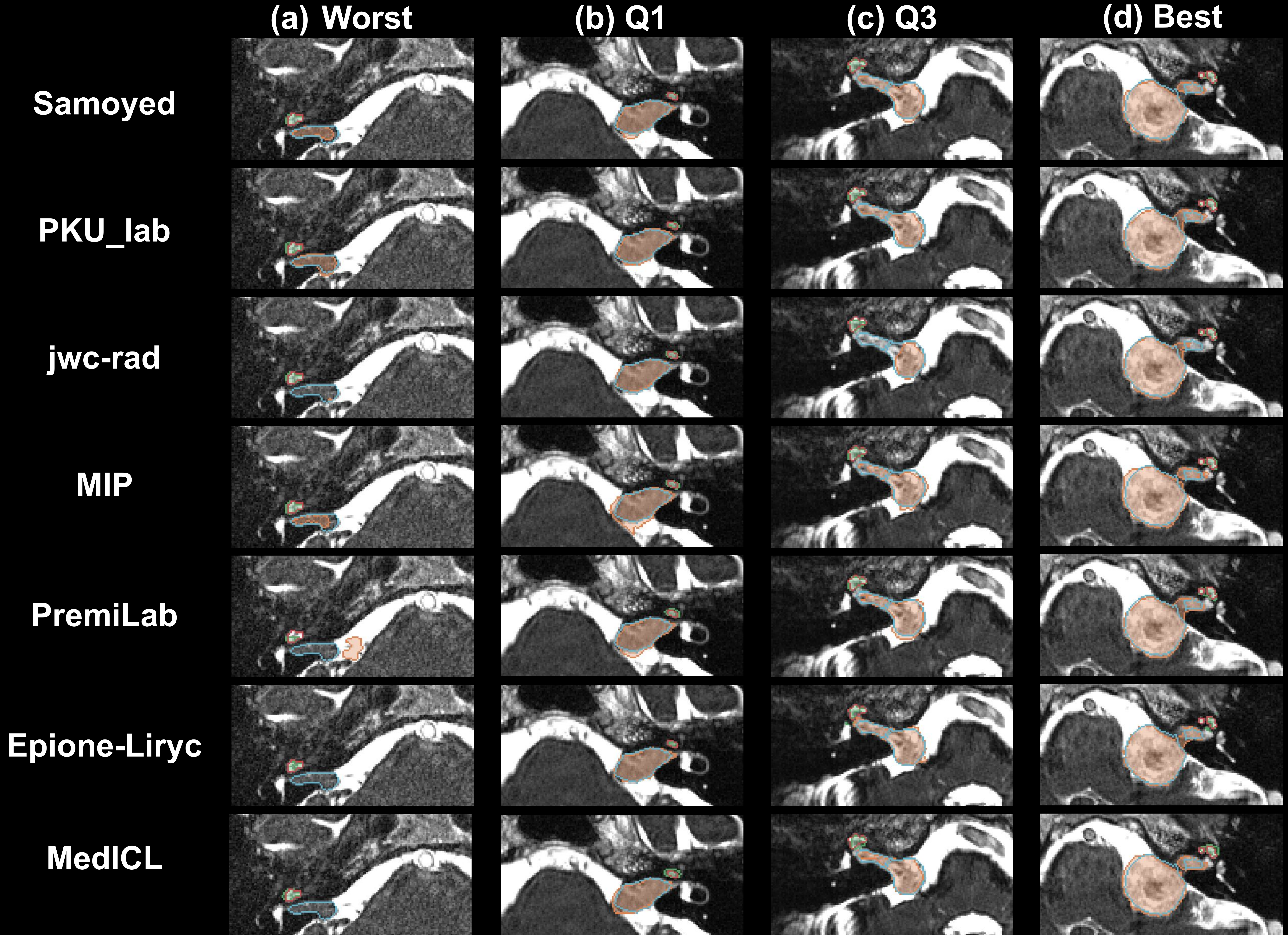}
  \caption{Qualitative comparison of the top 7 performing teams. Selected cases correspond to the (a) lowest, (b) lower-quartile, (c) upper-quartile, and (d) highest mean Dice score (averaged over the top 7 team and over the two structures).}
  \label{fig:qualitative_comparision}
\end{figure*}

\section{Results}\label{sec:results}

Participants submissions were required to submit their Docker container by 15th August 2021. Winners were announced during the crossMoDA event at the MICCAI 2021 conference. This section presents the results obtained by the participant teams on the test set and analyses the stability and robustness of the proposed ranking scheme.

\subsection{Overall segmentation performance}
The final scores for the 16
teams are reported in Table~\ref{tab:Scores} in the order in which they ranked. Figures~\ref{fig:boxvs} and~\ref{fig:boxco} show the box-plots for each structure (VS and cochleas) and are colour-coded according to the team. The performance distribution is given for each metric (DSC and ASSD). Qualitative results are shown in Figure~\ref{fig:qualitative_comparision}.

The winner of the crossMoDA challenge is Samoyed with a rank score of 2.7. Samoyed is the only team that reached a median DSC greater than $85\%$ for both structures. Other teams in the top five also obtained outstanding results with a median DSC greater than $80\%$ for each structure. In contrast, 
the low DSC and ASSD scores of
the three teams with the lowest rank 
highlight the complexity of the cross-modality domain adaptation task.

The top ten teams all used an approach based on image-to-image translation. As shown in Table~\ref{tab:Scores}, the medians are significantly higher, and the interquartile ranges (IQRs) are smaller compared to other approaches. Approaches using MIND cross-modality features obtained the following places (eleventh and twelfth ranks), while those aiming at aligning the distribution of the features extracted from the source and target images obtained the last positions. This highlights the effectiveness of using CycleGAN and its extensions to bridge the gap between \Tone and \Ttwo scans. 

\subsection{Evaluation per structure and impact on the rank}
The level of robustness and performance of the proposed techniques highly depends on the structure, impacting the ranking.

Examining the distribution of the scores is crucial for analysing the robustness of the proposed methods. More variability can be observed in terms of algorithm performance for the tumour than for the cochleas. On average, the IQRs of the top 10 performing teams for the DSC and ASSD are respectively $2.6$ and $16$  times larger for VS than cochleas. Moreover, Figures~\ref{fig:boxco} and~\ref{fig:boxvs} show that there are more outliers for VS than for cochleas. This suggests that the proposed algorithms are less robust on VS than on cochleas. For example, the winning team obtained a relatively poor DSC ($<60\%$) for respectively $8\%$ (N=11) and $1\%$ (N=1) of the testing set on the tumour and the cochleas. This can be explained by the fact that cochleas are more uniform in terms of location, volume size and intensity distribution than tumours. This could also be the reason why techniques using feature alignment collapsed on VS task.

Conversely, the level of performance for the cochleas task had a stronger impact on the rank scores. Table~\ref{tab:Scores} shows that the top seven teams obtained a comparable performance on the VS task (median DSC - min: $83.3\%$; max: $88.4\%$), while more variability is observed on the cochleas task (median DSC - min: $75.2\%$; max: $85.7\%$). Table~\ref{tab:Rank_structure} shows the distribution of the individual cumulative ranks for each structure (VS and cochleas). It can be observed that the winner significantly outperformed all the other teams on the cochleas.
In contrast, while the second team obtained the best performance on the VS task (see Table~\ref{tab:Scores}), it didn't rank high enough on the cochleas task to win the challenge. Similarly, the fourth to seventh teams, which obtained comparable cumulative ranks on the VS task (median between 5 and 5.5), are ranked in the same order as their median cumulative ranks on the cochleas task. This shows that the performance of the top-performing algorithms on the cochleas was the most discriminative for the final ranking.

\begin{table}[tb]
	\centering
	\caption{Distribution of the individual cumulative ranks for each structure. Median and interquartile values are presented. 
	}\label{tab:Rank_structure}
	\resizebox{0.45\textwidth}{!}{
	\begin{tabular}{l c c c}
		\toprule
	    &  Challenge & Vestibular  \\ 
		& Rank & Schwannoma & \multirow{-2}{*}{Cochleas} \\
		\midrule

        \circle{Samoyed-c} Samoyed & 1 & 3.0 [2.0 - 4.5] & 1.0 [1.0 - 2.0]  \\ 
        
        \rowcolor{LightGray}
        \circle{PKU-BIALAB-c} PKU\textunderscore BIALAB & 2 & 1.5 [1.0 - 2.5] & 5.0 [3.5 - 6.0]  \\ 
        
        \circle{jwc-rad-c} jwc-rad & 3 & 5.0 [3.0 - 7.0] & 3.0 [2.0 - 4.5]  \\ 
        
        \rowcolor{LightGray}
        \circle{MIP-c} MIP & 4 & 5.0 [4.0 - 7.0] & 3.5 [2.0 - 4.5]  \\ 
        
        \circle{PremiLab-c} PremiLab & 5 & 5.5 [3.5 - 7.0] & 5.0 [3.5 - 6.0]  \\ 
        
        \rowcolor{LightGray}
        \circle{Epione-Liryc-c} Epione-Liryc & 6 & 5.0 [3.0 - 6.5] & 7.0 [6.5 - 8.0]  \\ 
        
        \circle{MedICL-c} MedICL & 7 & 5.0 [3.5 - 7.0] & 8.0 [7.0 - 9.0]  \\ 
        
        \rowcolor{LightGray}
        \circle{DBMI-pitt-c} DBMI\textunderscore pitt & 8 & 13.0 [10.5 - 13.5] & 5.0 [3.5 - 6.0]  \\ 
        
        \circle{Hi-Lib-c} Hi-Lib & 9 & 9.0 [7.5 - 10.5] & 8.5 [7.5 - 9.5]  \\ 
        
        \rowcolor{LightGray}
        \circle{smriti161096-c} smriti161096 & 10 & 8.0 [7.5 - 9.0] & 11.0 [10.5 - 12.5]  \\ 
        
        \circle{IMI-c} IMI & 11 & 10.0 [8.5 - 11.0] & 13.0 [12.0 - 14.5]  \\ 
        
        \rowcolor{LightGray}
        \circle{GapMIND-c} GapMIND & 12 & 11.0 [10.0 - 12.0] & 11.5 [11.0 - 13.0]  \\ 
        
        \circle{gabybaldeon-c} gabybaldeon & 13 & 11.0 [10.0 - 12.0] & 13.0 [12.0 - 14.0]  \\ 
        
        \rowcolor{LightGray}
        \circle{SEU-Chen-c} SEU\textunderscore Chen & 14 & 15.0 [14.5 - 15.5] & 11.5 [10.0 - 14.0]  \\ 
        
        \circle{skjp-c} skjp & 15 & 14.0 [13.0 - 15.0] & 15.0 [13.0 - 16.0]  \\ 
        
        \rowcolor{LightGray}
        \circle{IRA-c} IRA & 16 & 14.5 [14.0 - 15.5] & 15.0 [14.5 - 15.5]  \\

\bottomrule
	\end{tabular}
	}
\end{table}

\subsection{Remarks about the ranking stability}
It has been shown that challenge rankings can be sensitive to various design choices, such as the test set used for validation, the metrics chosen for assessing the algorithms’ performance and the scheme used to aggregate the values \citep{Maier-Hein2018}. In this section, we analyse and visualise the ranking stability with respect to these design choices.

\begin{figure}[tb!]
  \includegraphics[width=\linewidth]{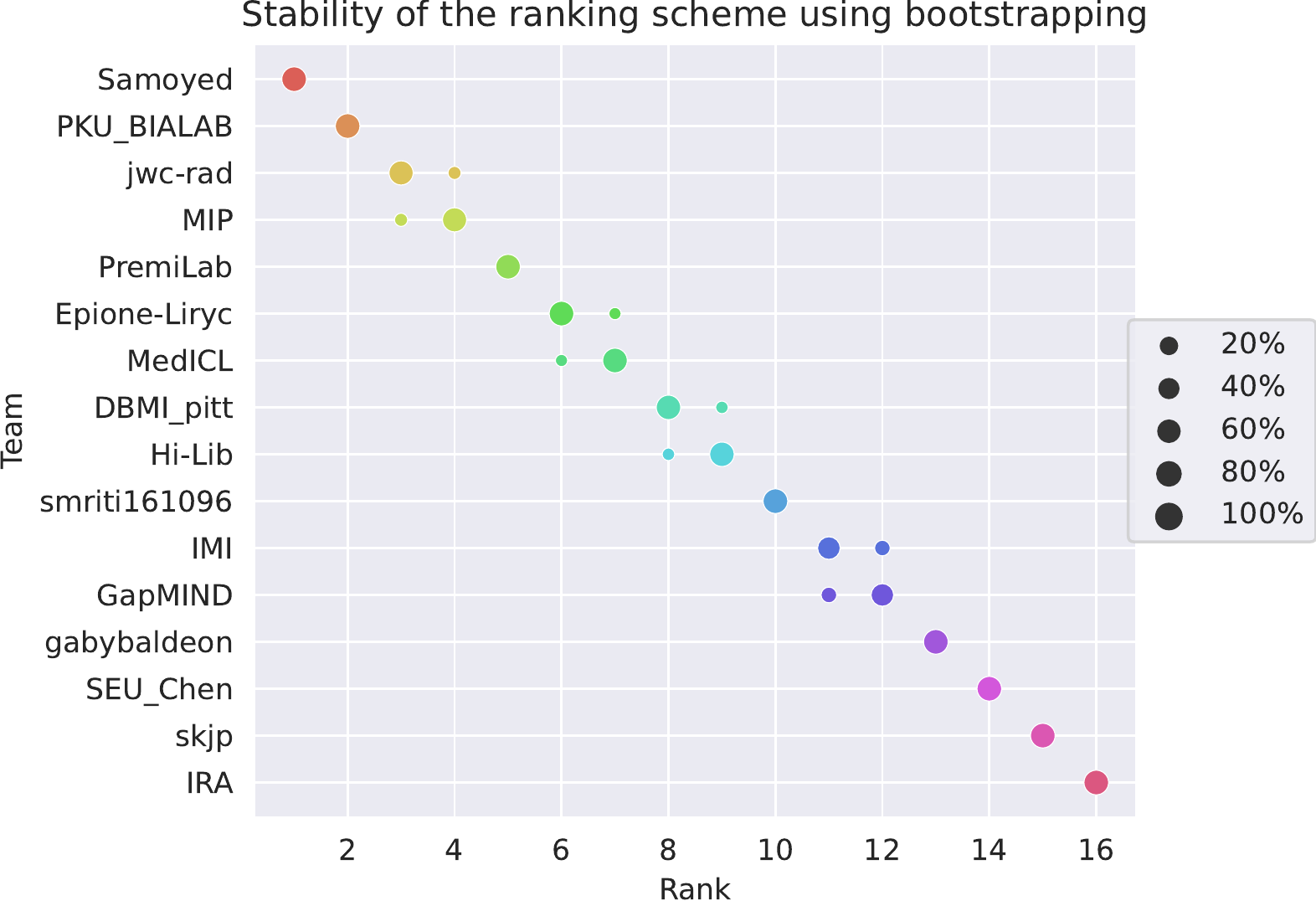}
  \caption{Stability of the proposed ranking scheme for 1000 bootstrap samples.}
  \label{fig:bootstrap_stability}
\end{figure}

A recent work proposed techniques to assess the stability of rankings with respect to sampling variability \citep{Wiesenfarth2021}. Following their recommendations, we performed bootstrapping (1,000 bootstrap samples) to investigate the ranking uncertainty and stability of the proposed ranking scheme with respect to sampling variability. To this end, the ranking strategy is performed repeatedly on each bootstrap sample. To quantitatively assess the ranking stability, the agreement of the challenge ranking and the ranking lists based on the individual bootstrap samples was determined via Kendall’s $\tau$, which provides values between $-1$ (for reverse ranking order) and $1$ (for identical ranking order). The median [IQR] Kendall’s $\tau$ was 1 [1-1], demonstrating the perfect stability of the ranking scheme. Figure~\ref{fig:bootstrap_stability} shows a blob plot of the bootstrap rankings. The same conclusion can be drawn: the ranking stability of the proposed scheme is excellent. In particular, the winning team is first-ranked for all the bootstrap samples.

\begin{figure}[tb!]
  \includegraphics[width=\linewidth]{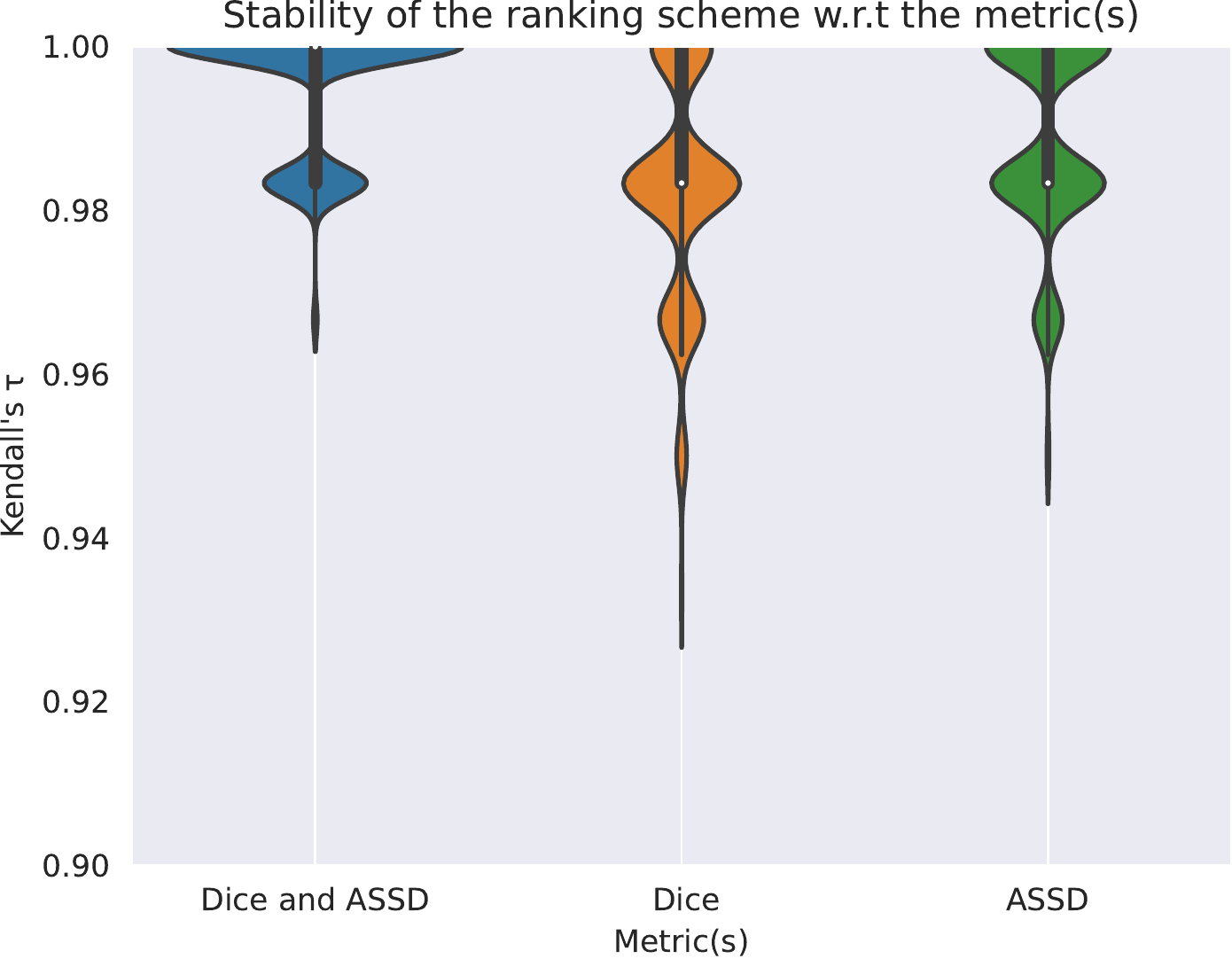}
  \caption{Stability of the ranking scheme with respect to the choice of the metrics. 1000 bootstrap samples are used.}
  \label{fig:metric_stability}
\end{figure}

To evaluate the stability of the ranking with respect to the choice of the metrics, we compared the stability of single-metric (DSC or ASSD) ranking schemes with our multi-metric (DSC and ASSD) ranking scheme.
Specifically, bootstrapping was used to compare the stability of the ranking for the three sets of metrics and Kendall’s $\tau$ were computed to compare the ranking list computed on the full assessment data and the individual bootstrap samples. Violin plots shown in Figure~\ref{fig:metric_stability} illustrate bootstrap results for each metric. It can be observed that Kendall’s $\tau$ are more dispersed across the bootstrap samples when using only one metric. Median Kendall’s $\tau$ are respectively $0.98$, $0.98$ and $1$ using DSC, ASSD and the combination of both as metric. This demonstrates that the ranking stability is higher when multiple metrics are used.

\begin{figure}[tb!]
  \includegraphics[width=\linewidth]{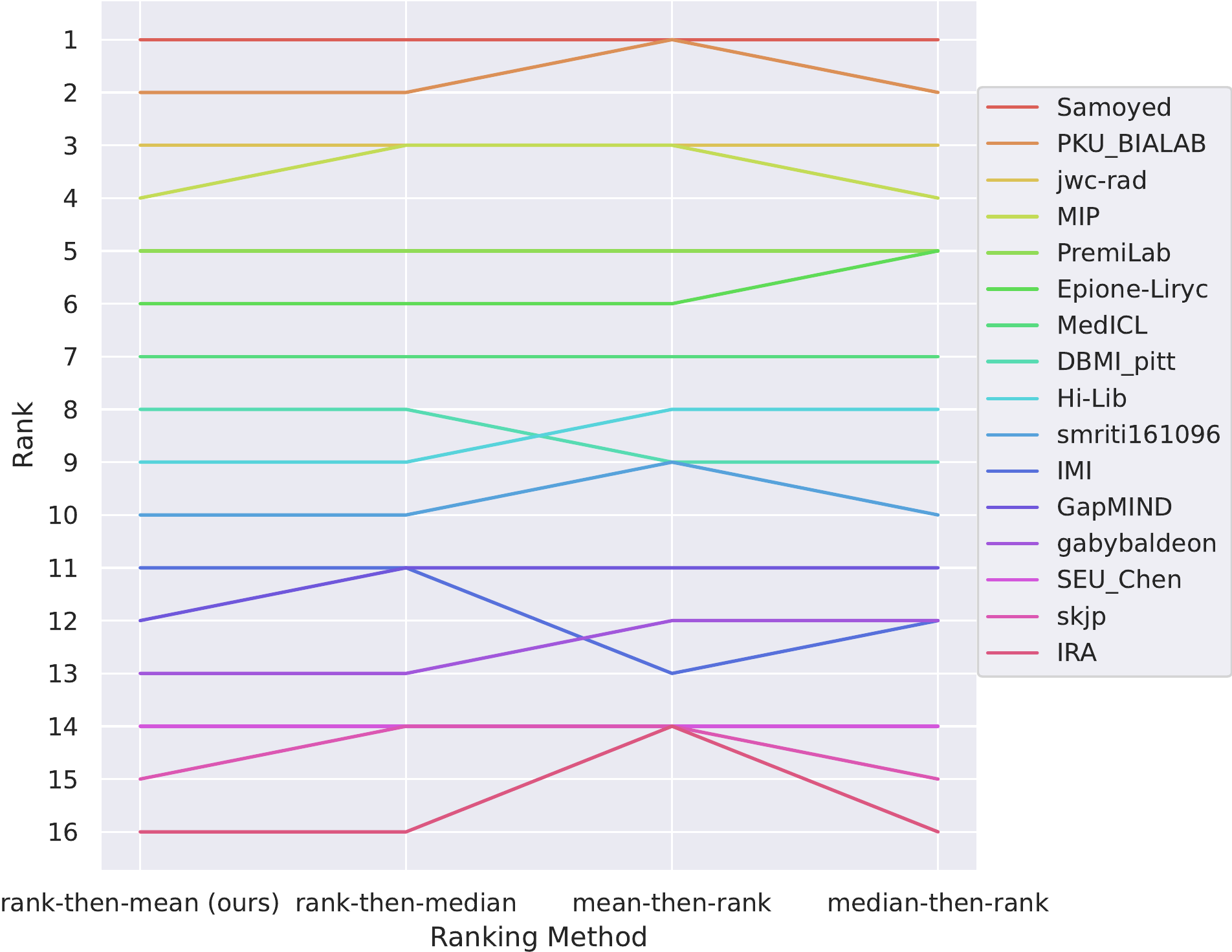}
  \caption{Line plots visualising rankings robustness across different ranking methods for the brain task. Each algorithm is represented by one coloured line. For each ranking method encoded on the x-axis, the height of the line represents the corresponding rank. The lowest rank of tied values is used for ties (equal scores for different teams) }
  \label{fig:ranking_methods}
\end{figure}

Finally, we compared our ranking scheme with other ranking methods with different aggregation methods. The most prevalent approaches are:
\begin{itemize}
    \item Aggregate-then-rank: metric values across all test cases are first aggregated (e.g., with the mean, median) for each structure and each metric. Ranks per structure and per metric are then computed for each team. Ranking scores correspond to the aggregation (e.g., with mean, median) of these ranks and are used for the final ranking.
    \item Rank-then-aggregate: algorithms’ ranks are computed for each test case, for each metric and each structure and then aggregated (e.g., with the mean, median). Then, the aggregated rank score is used to rank algorithms.
\end{itemize}
Our ranking scheme corresponds to a rank-then-aggregate approach with the mean as aggregation technique. We compared our approach with: 1/  a rank-then-aggregate approach using another aggregation technique (the median); 2/ aggregate-then-rank approaches using either the mean and the median for metric aggregation. 
Ranking robustness across these different ranking methods is shown on the line plots in Figure~\ref{fig:ranking_methods}. It can be seen that the ranking is robust to these different ranking techniques. In particular, the first seven ranks are the same for all ranking scheme variations. Note that the aggregate-then-rank approach using the mean is less robust due to the presence of outliers for the ASSD metric caused by missing segmentation for a given structure. This demonstrates that the ranking of the challenge is stable and can be interpreted with confidence.

\begin{table*}[tb]
	\centering
	\caption{ Impact of self-supervision on top 2 team methods. Median and interquartile values are presented. The best results are given in bold. Arrows indicate the favourable direction of each metric.
	}\label{tab:ablation_study}
	\begin{tabular}{l c *{5}{c}}
		\toprule
		\multirow{2}{*}{} & \multirow{2}{*}{\bf Self-Supervision } & \multicolumn{2}{c}{\bf Vestibular Schwannoma } & \multicolumn{2}{c}{\bf Cochleas }\\  
		\cmidrule(lr){3-4} \cmidrule(lr){5-6}
       & & DSC $(\%)$ $\uparrow$ & ASSD (mm) $\downarrow$ & DSC $(\%)$  $\uparrow$ & ASSD (mm) $\downarrow$ \\
		\midrule
          & $\checkmark$ & \textbf{87.0 [81.8 - 90.0]} & \textbf{0.39 [0.29 - 0.52]} & \textbf{85.7 [83.9 - 87.0]} & \textbf{0.13 [0.12 - 0.17]} \\ 
        
        \multirow{-2}{*}{\circle{Samoyed-c}Samoyed}  & $\times$ & 85.3 [79.7 - 89.1] & 0.42 [0.33 - 0.59] & 84.7 [81.9 - 86.3] & 0.15 [0.12 - 0.2] \\ 
        
        \rowcolor{LightGray}
          & $\checkmark$ & \textbf{88.4 [84.5 - 91.7]} & \textbf{0.31 [0.23 - 0.4]} & \textbf{80.6 [78.0 - 82.5]} & \textbf{0.18 [0.15 - 0.22]} \\ 
        
        \rowcolor{LightGray}
        \multirow{-2}{*}{\circle{PKU-BIALAB-c} PKU\textunderscore BIALAB}  & $\times$ & 86.5 [80.6 - 90.4] & 0.36 [0.27 - 0.5] & 75.7 [71.4 - 78.8] & 0.22 [0.18 - 0.28] \\ 
      
		\bottomrule
	\end{tabular}
\end{table*}

\section{Discussion}\label{sec:discussion}
In this study, we introduced the crossMoDA challenge in terms of experimental design, evaluation strategy, proposed methods and final results. In this section, we discuss the main insights and limitations of the challenge.

\subsection{Performance of automated segmentation methods}
To compare the level of performance reached by the top-performing teams with a fully-supervised approach, we trained a nnU-Net framework \citep{isensee2021nnu} on the \Ttwo scans paired with the \Tone scans from the source training set ($N=105$) and their manual annotations. Segmentation performances on VS and cochleas are reported in Table~\ref{tab:Scores}. It can be observed that full supervision significantly outperforms all participating teams on the two structures. The performance gap between the best performing team on VS and the fully-supervised model is $4.1\%$ (median DSC) and $0.11\text{mm}$ (median ASSD). The performance gap between the best performing team on cochleas and the fully-supervised model is $2\%$ (median DSC) and $0.03\text{mm}$ (median ASSD). Moreover, a fully-supervised approach obtained tighter IQRs, demonstrating better robustness. This shows that even though top-performing teams obtained a high level of performance, full supervision still outperforms these proposed approaches. Note that the top performing teams reached a level of performance that is higher than the one reported in another study with a weakly-supervised approach trained using scribbles on the VS and on a different split of the dataset \cite{ScribDA2020Dorent}.

\subsection{Analysis of the top-ranked methods}
Image-to-image translation using CycleGAN and its extension was the most successful approach to bridge the gap between the source and target images. Except for one team, 2D image-to-image translation was performed on 2D axial slices. Teams used CycleGAN and two of its extensions (NiceGAN, CUT).

However, the
results in Table~\ref{tab:Scores} do not demonstrate the advantage of using one image-to-image translation approach over another. For example, the third and ninth teams, which used the same implementation of the same approach (2D CUT), respectively obtained a median DSC of $83.1\%$ and $71.7\%$ on the cochleas. This shows that the segmentation performance depends on other parameters, such as the pre-processing step (cropping, image resampling, image normalisation) and the segmentation network. However, the current study does not allow for the identification of the optimal combination of parameters.

Except for three teams, all teams have used U-Net as segmentation backbone. In particular, the top three teams used the nnU-Net framework, a deep segmentation method that automatically configures itself, including pre-processing, network architecture, training and post-processing based on heuristic rules. This suggests the effectiveness of this framework for VS and cochleas segmentation.

Finally, it can be seen that the two top-performing teams used self-training. To study the impact of self-supervision on the framework performance, an ablation study was performed. Specifically, the two top-performing teams were asked to train their framework without the self-supervision component. Results are shown in Table~\ref{tab:ablation_study}. It can be seen that self-supervision leads to statically significant performance improvement provided by a Wilcoxon test ($p < 0.01$) for both teams. Self-training for domain adaptation has been previously proposed for medical segmentation problems. However, it is the first time that self-training has been successfully used in the context of large domain gaps. In practice,  image-to-image translation and self-training are used sequentially: first pseudo-target images are generated and used to train segmentation networks, and then self-training is used to fine-tune the trained networks to manage the (small) domain gap between pseudo- and real target images.

\subsection{Limitations and future directions for the challenge}
The lack of robustness to unseen situations is a key problem for deep learning algorithms in clinical practice. We created this challenge to benchmark new and existing domain adaptation techniques on a large and multi-class dataset. In this challenge, the domain gap between the source and target images is large, as it corresponds to different imaging modalities. However, the lack of robustness can also occur when the same image modalities are acquired in different settings (e.g., hospital, scanner). This problem is not addressed in this challenge. Indeed, images within a domain (target or source) have been acquired at a unique medical centre with the same scanner. Moreover, $96\%$ of the test set has been acquired with the same sequence parameters. For this reason, we plan to diversify the challenge dataset by adding data from other institutions. In particular, different \Ttwo appearances are likely to occur, making it challenging for image-to-image translation approaches, which assume that the relationship between the target and source domains is a bijection.

\section{Conclusion}\label{sec:conclusion}
The crossMoDA challenge was introduced to propose the first benchmark of domain adaptation techniques for medical image segmentation. The level of performance reached by the top-performing teams is surprisingly high and close to full supervision. Top performing teams all used
an image-to-image translation
approach to transform the source images into pseudo-target images and then train a segmentation network using these generated images and their manual annotations. Self-training has been shown to lead to performance improvements.

\section*{CRediT authorship contribution statement}
\textbf{Reuben Dorent:} Conceptualization, Methodology, Software, Formal analysis, Resources, Data curation, Writing - original draft, Writing - review $\&$ editing, Visualization. 
\textbf{Aaron Kujawa:} Conceptualization, Methodology, Data curation, Writing - review $\&$ editing.
\textbf{Spyridon Bakas:} Conceptualization, Methodology.
\textbf{Nicola Rieke:} Conceptualization, Methodology.
\textbf{Samuel Joutard:} Conceptualization, Methodology.
\textbf{Ben Glocker:} Conceptualization, Methodology.
\textbf{Jorge Cardoso:}  Conceptualization, Methodology.
\textbf{Marc Modat:} Conceptualization, Methodology.
\textbf{Kayhan Batmanghelich:} Methodology, Software.
\textbf{Arseniy Belkov:} Methodology, Software.
\textbf{Maria Baldeon Calisto:} Methodology, Software.
\textbf{Jae Won Choi:} Methodology, Software.
\textbf{Benoit M. Dawant:} Methodology, Software.
\textbf{Hexin Dong:} Methodology, Software.
\textbf{Sergio Escalera:} Methodology, Software.
\textbf{Yubo Fan:} Methodology, Software.
\textbf{Lasse Hansen:} Methodology, Software.
\textbf{Mattias P. Heinrich:} Methodology, Software.
\textbf{Smriti Joshi:} Methodology, Software.
\textbf{Victoriya Kashtanova:} Methodology, Software.
\textbf{Hyeon gyu Kim:} Methodology, Software.
\textbf{Satoshi Kondo:} Methodology, Software.
\textbf{Christian N. Kruse:} Methodology, Software.
\textbf{Susana K. Lai-Yuen:} Methodology, Software.
\textbf{Hao Li:} Methodology, Software.
\textbf{Han Liu:} Methodology, Software.
\textbf{Buntheng Ly:} Methodology, Software.
\textbf{Ipek Oguz:} Methodology, Software.
\textbf{Hyungseob Shin:} Methodology, Software.
\textbf{Boris Shirokikh:} Methodology, Software.
\textbf{Zixian Su:} Methodology, Software.
\textbf{Guotai Wang:} Methodology, Software.
\textbf{Jianghao Wu:} Methodology, Software.
\textbf{Yanwu Xul:} Methodology, Software.
\textbf{Kai Yao:} Methodology, Software.
\textbf{Li Zhang:} Methodology, Software.
\textbf{Sébastien Ourselin:} Supervision, Funding acquisition.
\textbf{Jonathan Shapey:} Conceptualization, Methodology, Data curation, Writing - review $\&$ editing, Funding acquisition.
\textbf{Tom Vercauteren:} Project administration, Conceptualization, Methodology, Formal analysis, Resources, Writing - original draft, Writing - review $\&$ editing, Funding acquisition.

\section*{Acknowledgements}
We would like to thank all the other team members that helped during the challenge:
Sewon Kim, Yohan Jun, Taejoon Eo, Dosik Hwang (Samoyed); 
Fei Yu, Jie Zhao, Bin Dong (PKU\textunderscore BIALAB); Can Cui, Dingjie Su, Andrew Mcneil (MIP);
Xi Yang, Kaizhu Huang, Jie Sun (PremiLab);
Yingyu Yang, Aurelien Maillot, Marta Nunez-Garcia, Maxime Sermesant (Epione-Liryc);
Dewei Hu, Qibang Zhu, Kathleen E Larson, Huahong Zhang (MedICL);
Mingming Gong (DBMI\textunderscore pitt);
Ran Gu, Shuwei Zhai, Wenhui Lei (Hi-Lib);
Richard Osuala, Carlos Martın-Isla, Victor M. Campello, Carla Sendra-Balcells, Karim Lekadir (smriti161096);
Mikhail Belyaev (IRA).

This work was supported by the Engineering and Physical Sciences Research Council (EPSRC) [NS/A000049/1, NS/A000050/1], MRC (MC/PC/180520) and Wellcome Trust [203145Z/16/Z, 203148/Z/16/Z, WT106882]. TV is supported by a Medtronic / Royal Academy of Engineering Research Chair [RCSRF1819\textbackslash7\textbackslash34]. 
Z.S and K.Y. are supported by the National Natural Science Foundation of China [No. 61876155], the Jiangsu Science and Technology Programme (Natural Science Foundation of Jiangsu Province) [No. BE2020006-4] and the Key Program Special Fund in Xi'an Jiaotong-Liverpool University (XJTLU) [KSF-E-37].
C.K. and M.H. are supported by the Federal Ministry of Education and Research [No. 031L0202B].
H.S. and H.G.K. are supported by Basic Science Research Program through the National Research Foundation of Korea (NRF) funded by the Ministry of Science and ICT [2019R1A2B5B01070488, 2021R1A4A1031437], Brain Research Program through the NRF funded by the Ministry of Science, ICT $\&$ Future Planning [2018M3C7A1024734], Y-BASE $R\&E$ Institute a Brain Korea 21, Yonsei University, and the Artificial Intelligence Graduate School Program, Yonsei University [No. 2020-0-01361].
H.Liu, Y.F. and B.D. are supported by the National Institute of Health (NIH) [R01 DC014462].
L.H. and M.H. are supported by the German Research Foundation (DFG) under grant number 320997906 [HE 7364/2-1].
S.J. and S.E. are supported by the Spanish project PID2019-105093GB-I00 and by ICREA under the ICREA Academia programme
B.L. and V.K. are supported by the French Government, through the National Research Agency (ANR) 3IA Côte d’Azur [ANR-19-P3IA-0002], IHU Liryc [ANR- 10-IAHU-04]. The Epione-Liryc team is grateful to the OPAL infrastructure from Université Côte d'Azur for providing resources and support.
H.Li and I.O are supported by the National Institute of Health (NIH) [R01-NS094456].
L.Z. and H.D. are supported by the Natural Science Foundation of China (NSFC) under Grants 81801778, 12090022, 11831002.
Y.X. and K.B. are supported by NIH Award Number 1R01HL141813-01, NSF 1839332 Tripod+X, SAP SE, and Pennsylvania's Department of Health and are grateful for the computational resources provided by Pittsburgh Super Computing grant number TG-ASC170024.
S.B. is supported by the National Cancer Institute (NCI) and the National Institute of Neurological Disorders and Stroke (NINDS) of the National Institutes of Health (NIH), under award numbers NCI:U01CA242871 and NINDS:R01NS042645. The content of this publication is solely the responsibility of the authors and does not represent the official views of the NIH.





\bibliographystyle{model2-names.bst}\biboptions{authoryear}
\bibliography{refs}

\end{document}